\documentclass[%
 preprint,
 amsmath,amssymb,
 aps,
]{revtex4-2}
\usepackage{graphicx}
\usepackage{dcolumn}
\usepackage{bm}

\newcommand{\etal}{\mbox{\textit{et al.\ }}}
\newcommand{\rec}{\mbox{\textit{Re}}}
\newcommand{\bs}{\boldsymbol}
\newcommand{\be}{\begin{equation}}
\newcommand{\ee}{\end{equation}}
\newcommand{\bea}{\begin{eqnarray}}
\newcommand{\eea}{\end{eqnarray}}
\newcommand{\pd}{\partial}
\newcommand\veps{\varepsilon}
\newcommand{\bnabla}{\boldsymbol{\nabla}}
\newcommand{\shdt}{\!\cdot\!}
\newcommand{\tpr}{t^{\prime}}
\newcommand{\spr}{s^{\prime}}
\usepackage[dvipsnames]{xcolor}
\usepackage{todonotes}

\usepackage{color}

\begin{document}

\preprint{APS/123-QED}

\title{Hydrodynamic memory and Quincke rotation}

\author{Jason K. Kabarowski}
\affiliation{Department of Chemical Engineering, Carnegie Mellon University, Pittsburgh, PA 15213, USA}
\author{Aditya S. Khair}%
\email{akhair@andrew.cmu.edu}
\affiliation{Department of Chemical Engineering, Carnegie Mellon University, Pittsburgh, PA 15213, USA}
\author{Rahil N. Valani}
\affiliation{Rudolf Peierls Centre for Theoretical Physics, Parks Road,
University of Oxford, OX1 3PU, United Kingdom}

\date{\today}

\begin{abstract}
The spontaneous (so-called Quincke) rotation of an uncharged, solid, dielectric, spherical particle under a steady uniform electric field is analyzed, accounting for the inertia of the particle and the transient fluid inertia, or ``hydrodynamic memory,'' due to the unsteady Stokes flow around the particle.
The dynamics of the particle are encapsulated in three coupled nonlinear integro-differential equations for the evolution of the angular velocity of the particle, and the components of the induced dipole of the particle that are parallel and transverse to the applied field.
These equations represent a generalization of the celebrated Lorenz system.
A numerical solution of these `modified Lorenz equations' (MLE) shows that hydrodynamic memory leads to an increase in the threshold field strength for chaotic particle rotation, which is in qualitative agreement with experimental observations.
Furthermore, hydrodynamic memory leads to an increase in the range of field strengths where multi-stability between steady and chaotic rotation occurs.
At large field strengths, chaos ceases and the particle is predicted to execute periodic rotational motion.
\end{abstract}

\maketitle

\section{\label{sec:int}Introduction}
The spontaneous rotary motion of a solid, dielectric particle in a low conductivity fluid under a direct current (DC) field was observed by Quincke in 1896 and is now commonly referred to as `Quincke rotation' \cite{Quincke1896}. 
There has been a resurgence of interest in this phenomenon in recent years with applications in colloidal directed assembly \cite{Dolinsky2012, Sherman2020}, droplet electro-hydrodynamics \cite{Salipante2010, Ouriemi2015, Vlahovska2019}, and active matter \cite{Bricard2015, Das2019, Pradillo2019, Karani2019, Zhu2019}.
Quincke rotation has also been shown to alter suspension rheology \cite{Cebers2000, Cebers2004, Lemaire2008, Belovs2020} and conductivity \cite{Pannacci2007}. 
For a sphere or long circular cylinder the rotation occurs only if the charge relaxation time of the particle, $\tau_2=\veps_2/\sigma_2$, is greater than that of the fluid, $\tau_1=\veps_1/\sigma_1$, where $\veps$ is the permittivity and $\sigma$ is the electrical conductivity.
Here, we consider that the field is applied perpendicular to the long axis of a cylinder, for which rotation occurs about that axis.
(The subscripts 1 and 2 refer to the fluid and particle, respectively.)
In this case, the base state has no fluid or particle motion and the induced dipole vector $\bs{p}$ inside the particle is oriented anti-parallel to applied field.
Above a threshold field strength this mis-aligned dipole is unstable to infinitesimal perturbations and spontaneous rotation of the particle ensues, driven by an electric (Maxwell) torque arising from the induced dipole having a component transverse to the applied field. 
The angular velocity vector is orthogonal to the electric field; thus, a cylinder rotates about its axis of symmetry, and a sphere rotates about an arbitrary direction in the plane perpendicular to the field.
In contrast, if $\tau_2<\tau_1$ the base state corresponds to an induced dipole parallel to the field, for which the electric torque dampens perturbations; this base state is stable.

If the inertia of the particle and fluid are neglected, the angular momentum balance on the particle requires that the sum of electric and hydrodynamic torques is zero.
In the absence of inertia, the hydrodynamic torque, $\bs{L}_H$, is given by the quasi-steady Stokes equations: for a sphere $\bs{L}_H=-8\pi\mu a^3\bs{\Omega}$, and for a cylinder the torque per unit length $\bs{L}_H=-4\pi\mu a^2\bs{\Omega}$, where $\mu$ is the fluid viscosity; $a$ is the radius for a sphere and cross-sectional radius for a cylinder; and $\bs{\Omega}$ is the angular velocity of the object. 
Consequently, for an inertialess sphere or cylinder in an inertialess fluid it is predicted that there is a single supercritical pitchfork bifurcation corresponding to the transition from a stationary base state to steady rotation.
Note, the Quincke rotation dynamics of a sphere and cylinder are mathematically equivalent, as will be shown later.
In the absence of inertia, the critical field strengths for the onset of rotation for a long circular cylinder or sphere have the same functional form of $E_c=\sqrt{2\mu/[\veps_1\tau_{MW}(\veps_{21}-\sigma_{21})]}$, where for a sphere $\veps_{21}=(\veps_2-\veps_1)/(\veps_2+2\veps_1)$ and $\sigma_{21}=(\sigma_2-\sigma_1)/(\sigma_2+2\sigma_1)$, and for a cylinder $\veps_{21}=(\veps_2-\veps_1)/(\veps_2+\veps_1)$ and $\sigma_{21}=(\sigma_2-\sigma_1)/(\sigma_2+\sigma_1)$ \cite{Das-13}.
However, the numerical values for $E_c$ for a sphere and cylinder could be unequal due to the different definitions of $\veps_{21}$ and $\sigma_{21}$ in each case.
Here, $\tau_{MW}$ is the Maxwell-Wagner relaxation time: for a sphere $\tau_{MW}=(\veps_2+2\veps_1)/(\sigma_2+2\sigma_1)$, and for a cylinder $\tau_{MW}=(\veps_2+\veps_1)/(\sigma_2+\sigma_1)$ \cite{Jon-84}.
Since $E_c$ is a real quantity this means rotation is possible only if $\veps_{21}-\sigma_{21}>0$, which is equivalent to $\tau_2>\tau_1$.
The rotation rate of the particle is $\Omega=\sqrt{[(E_a/E_c)^2-1]}/\tau_{MW}$, where $E_a=|\bs{E}_a|$. Recall, the angular velocity is orthogonal to the applied field, $\bs{E}_a$, i.e. $\bs{E}_a\cdot\bs{\Omega}=0$.
Therefore, the particle undergoes a supercritical pitchfork bifurcation at an applied field strength $E_a=E_c$: for $E_a<E_c$ there is a stable equilibrium in which the particle is non-rotating with an anti-parallel dipole; for $E_a>E_c$ this state is unstable and two stable equilibria correspond to a particle rotating at $\pm\Omega$ with a dipole that has a non-zero component transverse to the applied field. 
Which of the stable equilibria is attained, $+\Omega$ or $-\Omega$, is dictated by chance.

Experiments by Lemaire and Lobry \cite{Lemaire2002} and Peters \etal \cite{Peters2005} observed, however, that the steady rotation is replaced by chaotic variations in the magnitude and direction of rotation above a second threshold field strength.
Those authors showed that inclusion of \textit{particle} inertia in the angular momentum balance leads to the Quincke rotation dynamics being described \textit{exactly} by the celebrated Lorenz equations \cite{Lorenz1963}.
Inclusion of particle inertia does not affect the predicted value of $E_c$ or the steady rotation rate after the first, or primary, bifurcation.
The Lorenz equations are well known to exhibit chaotic dynamics: in particular, they exhibit a secondary bifurcation from a non-trivial steady state to chaos, as observed in the Quincke rotor experiments \cite{Lemaire2002,Peters2005}.
However, the measured threshold field strength for the secondary bifurcation (steady to chaotic rotation) is larger than that predicted from the Lorenz equations.
Peters \etal \cite{Peters2005} suggested that one source of this disagreement is their neglect of fluid inertia.
In experiments the Reynolds number associated with the particle rotation, $\rec=a^2\Omega/\nu$, is small compared to unity.
Here, $\Omega=|\boldsymbol{\Omega}|$ is the magnitude of the angular velocity and $\nu$ is the kinematic viscosity of the fluid. 
Consequently, the hydrodynamic torque is a linear function of the angular velocity.
However, unsteady diffusion of fluid momentum arising from the angular acceleration of the particle could still be important. 
This means that the hydrodynamic torque is not an instantaneous function of angular velocity, as would be predicted from the quasi-steady Stokes equations.
Instead, the unsteady Stokes equations must be used to calculate the hydrodynamic torque, which is now a function of the history of the angular velocity.
Thereby, fluid inertia endows the particle with a \textit{hydrodynamic memory} of its rotational motion.

Hydrodynamic memory becomes important in Quincke rotation when the momentum diffusion time is comparable to the Maxwell-Wagner time for relaxation of the induced dipole. 
The momentum diffusion time is $\tau_d=a^2/\nu$.
Thus, the dimensionless group $\gamma=\tau_d/\tau_{MW}$ characterizes the relevance of hydrodynamic memory: neglecting hydrodynamic memory is valid only if $\gamma\ll 1$; physically, here momentum diffuses quickly compared to the dipole relaxation. 
This condition is usually met for colloidal scale particles but not for larger, millimeter-sized, particles that are encountered in experiments and applications involving Quincke rotation \cite{Lemaire2002,Peters2005,Dom-16,Bro-16,Roz-21}; here, hydrodynamic memory can be important since $\gamma$ is not small.
As an example, Peters \etal \cite{Peters2005} examined rotation of a glass capillary of $a=1\mathrm{mm}$ in transformer oil, for which they estimate $\tau_d=70\,\mathrm{ms}$ and $\tau_{MW}=150\,\mathrm{ms}$, yielding $\gamma=0.47$.
Here, inertial effects result in chaotic rotation at sufficiently large fields.
However, the mathematical model constructed by these authors included only the inertia of the particle, which is incomplete, as commented by themselves, for the following reason.
Balancing the angular acceleration of the particle, which scales as $I\Omega/\tau_p$, where $I$ is the moment of inertia ($I=2ma^2/5$ for a sphere of mass $m$), with the quasi-steady hydrodynamic torque, which for a sphere scales as $8\pi\mu a^3\Omega$, gives the characteristic time for relaxation of the particle inertia as $\tau_p=(1/15)(\rho_2/\rho_1)(a^2/\nu)$. 
This gives $\tau_p/\tau_d=(1/15)(\rho_2/\rho_1)$, implying that the particle inertia and fluid inertia time scales are comparable, unless the particle is very much heavier or lighter than the fluid.
Hence, one cannot accurately model inertial effects on Quincke rotation by only considering the inertia of the particle; fluid inertia via hydrodynamic memory must be accounted for.
Therefore, our goal is to quantify the impact of hydrodynamic memory on Quincke rotation.
To that end, we will derive a dynamical system for the rotor dynamics including hydrodynamic memory, in terms of the time evolution of the angular velocity of the particle and its dipole moment.
The inclusion of hydrodynamic memory means that this dynamical system is integro-differential: we will call this set of equations as the `memory Lorenz equations' (MLE), since they represent a generalization of the classic Lorenz system~\cite{Lorenz1963}.

In section \ref{sec:gov}, we derive the MLE, and then in section \ref{sec:smle} we derive an \textit{ad hoc} reduced form thereof that is amenable to analytical investigation, in which the memory kernel of the integro-differential term is simplified. In section \ref{sec:lin} we conduct a linear stability analysis of the trivial solution of the MLE, showing that the first bifurcation is unaffected by hydrodynamic memory. In section \ref{sec:lin steady} we present a linear stability of the steady rotation, which indicates that hydrodynamic memory leads to an increase in the threshold field strength for which this solution becomes unstable. In section \ref{sec:num}, we develop a numerical scheme for integration of the MLE. In section \ref{sec:res}, we explore the nonlinear dynamics beyond steady rotation and present results that show chaotic rotation is delayed until higher electric field strengths by hydrodynamic memory. We conclude in section \ref{sec:con}.

\section{Governing Equations}\label{sec:gov}

\begin{figure}
  \centerline{\includegraphics[scale = 1]{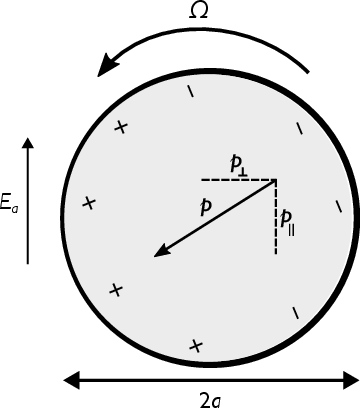}}
  \caption{Sketch of a spherical particle undergoing Quincke rotation. The induced dipole, $\boldsymbol{p}$, is misaligned with the electric field, $\boldsymbol{E}_a$, causing an electric torque to be exerted on the particle. This electric torque is balanced by a hydrodynamic torque, resulting in the sphere rotating with angular velocity, $\boldsymbol{\Omega}$. }
\label{fig:QuinckeDiagram}
\end{figure}

Here, we derive governing equations for the Quincke rotation of a solid spherical particle under a spatially uniform, DC electric field, accounting for hydrodynamic memory (figure \ref{fig:QuinckeDiagram}). 
We use the Taylor-Melcher leaky dielectric model \cite{Mel-69} in which charge transport is solely via Ohmic conduction in the bulk phases (particle and fluid), with charge localized at their interface.
Let $\bs{r}$ denote the position vector from the centroid of the sphere and let $\phi(\bs{r})$ denote the electric potential. 
The leaky dielectric model stipulates that there are no free charges in the fluid and particle; hence, the electric potential satisfies Laplace's equation, $\nabla^2\phi=0$.
The applied field corresponds to the dipolar potential $-\bs{E}_a\!\cdot\!\bs{r}$.
We define the disturbance potential in the fluid as $\phi_1(\bs{r})=\phi(\bs{r})+\bs{E}_a\!\cdot\!\bs{r}$ for $r=|\bs{r}|>a$ and the disturbance potential in the particle as $\phi_2(\bs{r})=\phi(\bs{r})+\bs{E}_a\!\cdot\!\bs{r}$ for $r=|\bs{r}|<a$.
These disturbance potentials arise as the particle and fluid have different electrical properties, such that the applied field is distorted by the particle.
The disturbance potentials satisfy $\nabla^2\phi_1=0$ and $\nabla^2\phi_2=0$, and $\phi_1$ decays as $r\to\infty$ and $\phi_2$ is bounded at $r=0$.
The free charge is localized at the fluid-particle interface and represented by a surface charge distribution $q(\bs{r})$ at $r=a$.
From Gauss' law the surface charge is equated to the jump in dielectric displacement across the interface, $q(\bs{r})=-\bs{n}\!\cdot\!(\veps_1\bnabla\phi_1-\veps_2\bnabla\phi_2)+\bs{n}\!\cdot\!\bs{E}_a(\veps_1-\veps_2)$.
Here, $\bs{n}$ is the outward unit vector to the particle surface.
Additionally, the potential is continuous across the interface, $\phi_1(\bs{r})=\phi_2(\bs{r})$ at $r=a$.
The final boundary condition enforces charge conservation at the interface,
\be\label{chgcons}
\frac{\pd q}{\pd t}+\bnabla_s\!\cdot\!(q\bs{U})+\bs{n}\!\cdot\!\bs{J}=0,
\ee
where $\bs{J}=(\sigma_1-\sigma_2)\bs{n}\cdot\bs{E}_a-\bs{n}\cdot(\sigma_1\bnabla\phi_1-\sigma_2\bnabla\phi_2)$ is the jump in Ohmic current at $r=a$.
Charge is convected along the interface at velocity $\bs{U}=a\bs{\Omega}\times\bs{n}$, where $\bs{\Omega}$ is the angular velocity of the particle, and $\bnabla_s=(\bs{I}-\bs{n}\bs{n})\!\cdot\!\bnabla$ is the surface gradient operator with $\bs{I}$ the identity tensor.
The value of $\bs{\Omega}$ is not known \textit{a priori} and must be solved for to determine the rotational dynamics.
For a sphere the disturbance potentials are dipolar solutions to Laplace's equation: $\phi_1=\bs{r}\shdt\bs{p}(t)/r^3$ and $\phi_2=\bs{r}\shdt\bs{p}(t)/a^3$, where $\bs{p}(t)$ is the time-dependent induced dipole vector.
These potentials satisfy the required boundary conditions at large and small $r$ and are continuous at $r=a$.
Using the expressions for $\phi_1$ and $\phi_2$ with the charge conservation condition \eqref{chgcons} yields an evolution equation for the dipole 
\be\label{diev}
\dot{\bs{p}}=\bs{\Omega}\times(\bs{p}-\veps_{21}a^3\bs{E})-\frac{1}{\tau_{MW}}(\bs{p}-\sigma_{21}a^3\bs{E}),
\ee
where $\dot{\bs{p}}=d\bs{p}/dt$.
The angular velocity $\bs{\Omega}$ is a function of $\bs{p}$ and thus \eqref{diev} is a nonlinear equation.

Next, we invoke angular momentum conservation on the particle:
\be\label{angmom}
I\frac{d\bs{\Omega}}{dt}=\bs{L}^E+\bs{L}^H,
\ee
which states that the angular acceleration (i.e., particle inertia) is balanced by the sum of electric and hydrodynamic torques.
The electric torque arises from Maxwell stresses and for a uniform applied field is $\bs{L}^E=4\pi\veps_1\bs{p}\times\bs{E}_a$ \cite{Was-94}.
This torque is non-zero if the dipole has a component that is transverse to the applied field.
The hydrodynamic torque is \cite{Pre-19}
\be\label{torhy}
\bs{L}^H=-8\pi\mu a^3\left[\bs{\Omega}+\int_{-\infty}^t\dot{\bs{\Omega}}(\tpr)M(t-\tpr)\,\mathrm{d}\tpr\right],
\ee 
where $\dot{\bs{\Omega}}(t^{\prime})$ denotes $d\bs{\Omega}(\tpr)/d\tpr$.
The first term in \eqref{torhy} is the quasi-steady torque; the second (integral) term arises from hydrodynamic memory, with the kernel
\be\label{memker}
M(t)=\frac{1}{3}\left[\sqrt{\frac{\tau_d}{\pi t}}-\mathrm{erfc}\left(\sqrt{\frac{t}{\tau_d}}\right)\exp\left(\frac{t}{\tau_d}\right)\right]\,\,\mathrm{for}\,\,t\geq0,
\ee
and $M(t)=0$ for $t<0$ (due to causality).
Here $\mathrm{erfc}(z)=\textstyle{\frac{2}{\sqrt{\pi}}}\int_z^{\infty}\exp(-u^2)\,\mathrm{d}u$ is the complimentary error function. 
Note that $M(t)= 1/[3\sqrt{\pi t/\tau_d}]+O(1)$ as $t\to0$ and $M(t)= (t/\tau_d)^{-3/2}/(6\sqrt{\pi})+O(t^{-5/2})$ as $t\to\infty$.
Thus, the memory kernel is integrably singular at short times, which will require the use of special quadrature schemes.
Together, \eqref{diev} through \eqref{memker} complete the formulation for Quincke rotation of a sphere with hydrodynamic memory.
Evidently, the history of the angular acceleration must be accounted for, which means that the rotational dynamics obey a coupled set of nonlinear integro-differential equations.

To proceed, we adopt a Cartesian coordinate system anchored at the center of the particle, with the $z$ axis parallel to the applied field and the $x$ and $y$ axes in the plane perpendicular to the field. 
There are components of the dipole along each of these axes, say $p_x$, $p_y$, and $p_z$.
From \eqref{angmom} we find $\bs{\Omega}\cdot\bs{E}_a=0$, so that the angular velocity only has non-zero components $\Omega_x$ and $\Omega_y$.
Thus, it would seem that the rotational dynamics obey five coupled ordinary differential equations for the variables $p_x$, $p_y$, $p_z$, $\Omega_x$, and $\Omega_y$, as a function of time.
However, these equations are invariant under the transformation $\{p_x,p_y,\Omega_x,\Omega_y\}\to\{p_y,p_x,-\Omega_y,-\Omega_x\}$, which suggests a solution with $p_x=p_y=p_{\perp}$ and $\Omega_x=-\Omega_y=-\Omega_{\perp}$.
This means that the direction of rotation in the $x-y$ plane is arbitrary, and the strength of the transverse dipole, $p_{\perp}$, is independent of orientation in the $x-y$ plane.
Thus, we have the dependent variables $p_z,p_{\perp}$, and $\Omega_{\perp}$, which from \eqref{diev} and \eqref{angmom} satisfy
 \bea
 \dot{p}_z&=&2\Omega_{\perp}p_{\perp}-\frac{1}{\tau_{MW}}(p_z-\sigma_{21}a^3E_a),\label{pzdim}\\
 \dot{p}_{\perp}&=&-\Omega_{\perp}p_z+\veps_{21}a^3E_a\Omega_{\perp}--\frac{1}{\tau_{MW}}p_{\perp},\label{pperdim}\\
 I\dot{\Omega}_{\perp}&=&4\pi\veps_1E_ap_{\perp}-8\pi\mu a^3\Omega_{\perp}-8\pi\mu a^3\int_0^t\dot{\Omega}_{\perp}(\tpr)M(t-\tpr)\,\mathrm{d}\tpr\label{omdim}.
 \eea
 The lower limit of the integral in \eqref{omdim} is zero since the field is initiated at $t=0$.
 We non-dimensionalize \eqref{pzdim}-\eqref{omdim} using the normalizations $s=t/\tau_{MW}$, $X=\sqrt{2}\tau_{MW}\Omega_{\perp}$, $Y=\veps_1\tau_{MW}p_{\perp}E_a/\sqrt{2}\mu a^3$, and $Z=\tau_{MW}E_a(p_z-\sigma_{21}a^3E_a)/8\pi\mu a^3$, which returns the dimensionless system
\bea
\dot{X}&=&Pr\left(Y-X-\int_0^s\dot{X}(\spr)M(s-\spr)\,\mathrm{d}\spr\right),\label{xeq}\\
\dot{Y}&=&-XZ+rX-Y,\label{yeq}\\
\dot{Z}&=&XY-Z,\label{zeq}
\eea
wherein the time derivatives are with respect to $s$; $M(s)=\textstyle{\frac{1}{3}}[1/\sqrt{\pi s}-\mathrm{erfc}(\sqrt{s/\gamma})\exp(s/\gamma)]$ is the normalized memory kernel; $Pr=8\pi\mu a^3\tau_{MW}/I$; and $r=(E_a/E_c)^2$.
Recall, $\gamma=\tau_d/\tau_{MW}$ is the ratio of momentum diffusion time to Maxwell-Wagner relaxation time.
Hydrodynamic memory effects vanish as $\gamma\to 0$, in which case the integral term in \eqref{xeq} becomes negligible.
In fact, without that integral term \eqref{xeq}-\eqref{zeq} are precisely the celebrated Lorenz equations~\citep{Lorenz1963}.
The Lorenz equations also describe the Quincke rotation of a cylinder in an inertialess fluid: here, the normalized angular velocity $X$ is along the axis of the cylinder, and $Y$ and $Z$ are the transverse and parallel components of the dipole in the plane of the cross section of the cylinder \cite{Peters2005}.
Thus, the Quincke rotation of a sphere and cylinder are mathematically equivalent.
We refer to \eqref{xeq}-\eqref{zeq}, with the integral term retained, as the `memory Lorenz equations' (MLE).
The MLE account for the inertia of the particle and surrounding fluid, both of which are important under physically realistic conditions.

\section{Lorenz equations with simplified memory (sMLE)}\label{sec:smle}

To understand more generally the role of memory in affecting the bifurcations of the Lorenz equations, we introduce simplified memory Lorenz equations (sMLE) with the memory kernel $M(s)=\alpha\,\text{e}^{-s/\alpha}$. This simplified memory kernel may be of practical relevance, in addition to mathematical convenience. Specifically, the stress in a linear viscoelastic fluid is a memory integral of the rate of strain, which captures the physical effect that viscoelasticity, akin to inertia, endows a fluid with a memory of the history of its deformation.  For the popular Maxwell model the kernel in the memory integral decays exponentially in time. Thus, the unsteady torque on a spherical particle executing time dependent rotation in a Maxwell fluid should have an exponentially fading memory of the history of its angular velocity, akin to \eqref{torhy}. That is, the sMLE may be relevant to Quincke rotation in a linear viscoelastic fluid. Here, the memory is provided by viscoelasticity as opposed to fluid inertia. The dimensionless parameter $\alpha$ in the sMLE would physically correspond to the ratio of the viscoelastic relaxation time of the fluid to the Maxwell-Wagner relaxation time. That is, $\alpha$ plays an analogous role to $\gamma$ in the MLE.

The exponentially decaying memory kernel allows us to convert the infinite-dimensional dynamical system of the Lorenz equations with an integral term into a fourth order system of nonlinear ODEs. The integral term in \eqref{xeq} with our simplified memory kernel is given by
\begin{align*}
    I(s)&=\int_0^s\frac{\mathrm{d} {X}(\spr)}{\mathrm{d} s'}M(s-\spr)\,\mathrm{d}\spr \\
    &=\int_0^s\frac{\mathrm{d}}{\mathrm{d} s'}\left({X}(\spr)M(s-\spr)\right)\,\mathrm{d}\spr - \int_0^s{X}(\spr)\frac{\mathrm{d} M(s-\spr)}{\mathrm{d} s'}\,\mathrm{d}\spr \\
    &=\left[ X(s) M(0) - X(0) M(s)\right] - \frac{1}{\alpha} \int_0^s{X}(\spr) M(s-\spr)\,\mathrm{d}\spr \\
    &=\alpha X(s) - \frac{1}{\alpha} \int_0^s{X}(\spr) M(s-\spr)\,\mathrm{d}\spr.
\end{align*}
In above, we have assumed that $X(0)=0$~\footnote{Even if $X(0)\neq0$, the additional term $X(0)M(s)$ goes to zero exponential as $s\xrightarrow{}\infty$.}. Now if we denote the history integral term by $H(s)=\int_{0}^s X(\spr)M(s-\spr)\,\mathrm{d}\spr$
then we get the system of equations
\begin{align}
    \dot{X}&=\text{Pr}(Y-X)-\frac{\text{Pr}} {\alpha} (\alpha^2 X-H), \label{eq: x sMLE} \\
    \dot{H}&=\frac{1}{\alpha} \left( \alpha^2 X-H\right), \label{eq: h sMLE} \\
    \dot{Y}&=-Y+rX-XZ, \label{eq: y sMLE} \\
    \dot{Z}&=-Z+XY. \label{eq: z sMLE}
\end{align}
Here dots denote derivative with respect to $s$. In getting the above equations, we have differentiated $H(s)$ with respect to $s$ and used Leibnitz integral rule for differentiation under the integral sign. Further, these equations reduce to the standard Lorenz equations in the limit $\alpha \xrightarrow{} 0$.



\section{Linear Stability Analysis of a stationary particle}\label{sec:lin}
\subsection{MLE model}
The MLE has a trivial solution of $(X, Y, Z) = (0, 0, 0)$, corresponding to a stationary particle whose dipole is anti-parallel to the applied field. We examine the stability of this base state by introducing a small perturbation $(X, Y, Z) = (0, 0, 0) + \delta (X_1, Y_1, Z_1)$, where $0<\delta \ll 1$, into the MLE and neglecting any terms quadratic or higher powers in $\delta$. 
This yields the following initial value problem for $(X_1,Y_1,Z_1)$, 
\begin{eqnarray}
    \frac{\mathrm{d} X_1}{\mathrm{d} s} &=& Pr \left( Y_1 - X_1 - \int_0^s \frac{\mathrm{d} X_1(s')}{\mathrm{d} s'}M(s - s') \mathrm{d} s' \right), \label{eqn:linearX} \\
    \frac{\mathrm{d} Y_1}{\mathrm{d} s} &=& rX_1 - Y_1, \label{eqn:linearY} \\
    \frac{\mathrm{d} Z_1}{\mathrm{d} s} &=& - Z_1 \label{eqn:linearZ},
\end{eqnarray}
which is solved via Laplace transform to yield
\begin{equation}
    \tilde{X}_1(\hat{s}) = \frac{X_1(0) + Pr\left( \tilde{M}(\hat{s}) X_1(0) - Y_1(0) (1 + \hat{s})^{-1}\right)}{\hat{s} - Pr\left( r (1 + \hat{s})^{-1} - 1 - \tilde{M}(\hat{s}) \hat{s}\right)},
    \label{eqn:linearstab}
\end{equation}
where an overhead $(\sim)$ denotes the Laplace transform, $\hat{s}$ is the Laplace frequency, and $\tilde{M}(\hat{s}) = \gamma / (3 [\sqrt{\gamma \hat{s}} + 1])$. While we are unable to invert (\ref{eqn:linearstab}) in closed form, progress is made at long times by taking the limit as $\hat{s} \rightarrow 0$ of $\tilde{X}_1(\hat{s})$, thereby yielding
\begin{equation}
    \lim_{\hat{s}\to 0} \tilde{X}_1(\hat{s}) = \frac{Pr\left(\frac{1}{3} \gamma X_1(0) - Y_1(0)\right) + X_1(0)}{\hat{s}-Pr\left(r - 1\right)}.
    \label{eqn:stabilitylongtime}
\end{equation}
Hence, for small $\hat{s}$, corresponding to long times, inversion of \eqref{eqn:stabilitylongtime} yields exponential solutions $X_1(s)\sim \exp[3Pr(r-1)s]$.
The initial perturbation grows~(decays) if the argument of the exponential is positive~(negative); the criterion for marginal stability is thus $r=1$, which in dimensional terms is $E_a=E_c$. A linear stability analysis of the standard Lorenz equations (LE) and the sMLE model (as shown below) also predicts the same criteria of instability of the stationary state. Therefore, hydrodynamic memory does not alter the threshold field strength for linear instability. Intuitively, this is expected since the base state is stationary.

\subsection{sMLE model}

The sMLE in \eqref{eq: x sMLE}-\eqref{eq: z sMLE} also have the trivial stationary solution $(X,Y,Z,H)=(0,0,0,0)$. We can understand the stability of this solution by again applying a small perturbation $(X,Y,Z,H)=(0,0,0,0)+\delta (X_1,Y_1,Z_1,H_1)$. Substituting this in \eqref{eq: x sMLE}-\eqref{eq: z sMLE}, we obtain the following linearized matrix equations for the evolution of perturbations

\begin{gather*}
 \begin{bmatrix} 
 \dot{X}_{1} \\
 \dot{Y}_1 \\
 \dot{Z}_1 \\
 \dot{H}_1 
 \end{bmatrix}
 =
  \begin{bmatrix}
-Pr(1+\alpha) & Pr & 0 & Pr/\alpha \\
r & -1 & 0 & 0 \\
0 & 0 & -1 & 0 \\
\alpha & 0 & 0 & -1/\alpha
 \end{bmatrix}
  \begin{bmatrix}
  X_{1}\\
  {Y}_1 \\
 {Z}_1 \\
 {H}_1 
 \end{bmatrix}.
\end{gather*}

The linear stability is determined by solving for the eigenvalues of the right-hand-side matrix. We get the following characteristic cubic polynomial for non-trivial eigenvalues $\lambda$ (in addition to the solution $\lambda=-1$)

\begin{equation}
 \alpha \lambda^3 + (1+\alpha+\alpha Pr + \alpha^2 Pr)\lambda^2 + (1+\alpha Pr + \alpha^2 Pr + (1-r \alpha) Pr) \lambda + (1-r) Pr =0.
\end{equation}

By using Descartes' rule of sign, we can deduce the existence of a positive root of this polynomial equation. The coefficients of $\lambda^3, \lambda^2$ and $\lambda$ are always positive whereas the constant term will be positive for $r<1$ and negative for $r>1$. For $r<1$, since all coefficients are positive, there are no sign changes and hence no positive roots exist. For $r>1$, there is one sign change between coefficients of consecutive terms in the polynomial and hence we can deduce the existence of one positive root. Thus, the instability threshold for the stationary state for the sMLE is also $r=1$.

\section{Linear Stability Analysis of steady rotation}\label{sec:lin steady}

\subsection{MLE model}

Once the stationary state becomes unstable for $r>1$, the system transitions to another steady state $(X, Y, Z) = (\pm \sqrt{r - 1}, \pm \sqrt{r - 1}, r - 1)$ corresponding to a steady angular velocity with clockwise or counterclockwise rotation. This steady state is also the same for LE and sMLE (as shown below). Thus, memory effects do not have an influence in setting the threshold for the primary pitchfork bifurcation as well as the corresponding steady state the system settles into after the bifurcation. Again, this is intuitive since at long times the particle is rotating steadily and the memory of its approach to that steady state fades as the memory kernel decays at large times. 
We examine the stability of the counterclockwise rotating steady state (without loss of generality) by introducing a small perturbation $(X, Y, Z) = ( \sqrt{r - 1}, \sqrt{r - 1}, r - 1) + \delta (X_1, Y_1, Z_1)$, where $0<\delta \ll 1$, into the MLE and neglecting any terms quadratic or higher in $\delta$. 
This yields the following initial value problem for $(X_1,Y_1,Z_1)$, 
\begin{eqnarray}
    \frac{\mathrm{d} X_1}{\mathrm{d} s} &=& Pr \left( Y_1 - X_1 - \int_0^s \frac{\mathrm{d} X_1(s')}{\mathrm{d} s'}M(s - s') \mathrm{d} s' \right), \label{eqn:linearX2} \\
    \frac{\mathrm{d} Y_1}{\mathrm{d} s} &=& rX_1 - Y_1 -\sqrt{r-1} 
 \,Z_1 - (r-1) X_1, \label{eqn:linearY2} \\
    \frac{\mathrm{d} Z_1}{\mathrm{d} s} &=& - Z_1 +\sqrt{r-1}\,X_1 + \sqrt{r-1}\,Y_1\label{eqn:linearZ2},
\end{eqnarray}

By applying Laplace transforms we get the following matrix equation for evolution of perturbations in the Laplace space
\begin{equation*}
\mathsf{A}(\hat{s})\mathbf{{X}}(\hat{s})=\mathbf{X}_0(\hat{s}),
\end{equation*}
where
\begin{gather*}
  \mathbf{X}(\hat{s})=
\begin{bmatrix}
    \tilde{X}_{1}(\hat{s})\\
    \tilde{Y}_{1}(\hat{s})\\
    \tilde{Z}_{1}(\hat{s})\\
\end{bmatrix},\,\,
  \mathbf{X}_0(\hat{s})=
\begin{bmatrix}
    {X}_{1}(0)+Pr \tilde{M}(\hat{s}) X_1(0)\\
    Y_1(0)\\
    Z_1(0)\\
\end{bmatrix},
\end{gather*}
and
\begin{gather*}
      \mathbf{A}(\hat{s})=
\begin{bmatrix}
    \hat{s} (1+Pr\,\tilde{M}(\hat{s}))+Pr & -Pr & 0\\
    -1 & \hat{s}+1 & \sqrt{r-1}\\
    -\sqrt{r-1} & -\sqrt{r-1} & \hat{s}+1\\
\end{bmatrix}.
\end{gather*}

The growth rates of this linear stability problem corresponds to the poles of $\mathbf{X}(\hat{s})$. Hence finding the growth rates reduces to determining the roots of $\det(\mathsf{A}(\hat{s}))=0$. We solve this numerically by finding solutions in the two dimensional space formed by real and imaginary part of $\hat{s}$ that satisfy both real and imaginary parts of $\det(\mathsf{A}(\hat{s}))=0$. The instability of steady rotation state takes place when the real part of $\hat{s}$ (that solves $\det(\mathsf{A}(\hat{s}))=0$) changes sign from negative to positive.

In figure \ref{fig:lin stab}(a) the marginal stability curves for the steady rotation branch in $r-Pr$ space are plotted for the MLE. It can be seen that an increase in memory (i.e. increasing $\gamma$) leads to a stabilization of the steady rotation state in the parameter space compared to the standard Lorenz equations (LE) shown as a gray curve. That is the critical value of $r$ for instability of the steady rotation increases with $\gamma$ for a fixed $Pr$.




\subsection{sMLE model}

Once the stationary state becomes unstable in the sMLE model, another steady state emerges given by $(X,Y,Z,H)=(\pm \sqrt{r - 1}, \pm \sqrt{r - 1}, r - 1,\pm \alpha^2 \sqrt{r - 1})$. This corresponds again to the steady rotation of the Quincke rotor. We performing a linear stability around the counterclockwise rotation state (without loss of generality) by applying a perturbation $(X,Y,Z,H)=(\sqrt{r - 1}, \sqrt{r - 1}, r - 1,\alpha^2 \sqrt{r - 1})+\delta (X_1,Y_1,Z_1,H_1)$, and substituting this in \eqref{eq: x sMLE}-\eqref{eq: z sMLE} we obtain the following linearized matrix equation

\begin{gather*}
 \begin{bmatrix} 
 \dot{X}_{1} \\
 \dot{Y}_1 \\
 \dot{Z}_1 \\
 \dot{H}_1 
 \end{bmatrix}
 =
  \begin{bmatrix}
-Pr(1+\alpha) & Pr & 0 & Pr/\alpha \\
1 & -1 & -\sqrt{r-1} & 0 \\
\sqrt{r-1} & \sqrt{r-1} & -1 & 0 \\
\alpha & 0 & 0 & -1/\alpha
 \end{bmatrix}
  \begin{bmatrix}
  X_{1}\\
  {Y}_1 \\
 {Z}_1 \\
 {H}_1 
 \end{bmatrix}.
\end{gather*}

The linear stability is determined by solving for the eigenvalues of the right-hand-side matrix equation. We get the following characteristic quartic polynomial

\begin{align}
 &\alpha \lambda^4 + (2\alpha+\alpha^2 Pr + 1 + Pr \alpha)\lambda^3 + (r \alpha+2\alpha^2 Pr + 2 + Pr \alpha + Pr)\lambda^2 \\ \nonumber
 &+ (r \alpha^2 Pr + r + 2 r Pr \alpha - 2 Pr \alpha + Pr) \lambda + 2(r-1) Pr =0.
\end{align}

We can numerically solve this quartic polynomial and find a boundary in the parameter space formed by $Pr$ and $r$ for a fixed $\alpha$ where an eigenvalue changes the sign of its real part.

\begin{figure}
  \centerline{\includegraphics[width = \columnwidth]{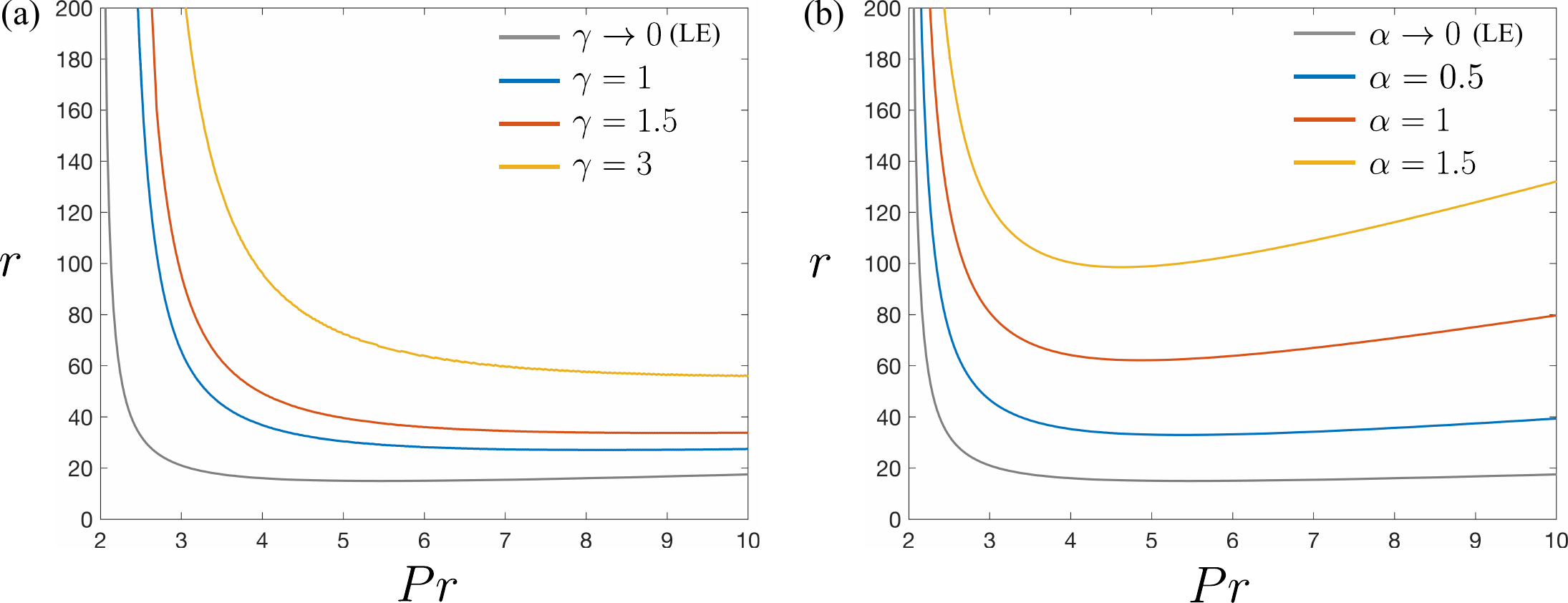}}
  \caption{Linear stability curves for steady rotation. Curves showing the linear stability boundary of the steady rotation branch in the parameter space of $Pr$ and $r$ for (a) MLE and (b) sMLE. The limit $\gamma \xrightarrow{} 0$ and $\alpha \xrightarrow{} 0$ corresponds to the Lorenz equations (LE), shown as a gray curve. This gray curve corresponding to instability of the steady state in LE takes the analytical form $r=Pr(Pr+4)/(Pr-2)$~\citep{Peters2005}.}
\label{fig:lin stab}
\end{figure}

In figure \ref{fig:lin stab}(b) the marginal stability curves for the steady rotation branch in $r-Pr$ space are plotted for sMLE. It can be seen that here too, an increase in memory (i.e. increasing $\alpha$) leads to a stabilization of the steady rotation state in the parameter space compared to LE.

\subsection{Beyond steady rotation}

For larger $r$ values beyond steady rotation, a complex set of bifurcations take place in LE~\cite{Sparrowbook,jackson_1990}. Furthermore, global bifurcations in LE take place even when the steady rotation state is stable resulting in coexistence of steady states and chaotic dynamics~\cite{Sparrowbook,jackson_1990}. Similar level of complexity would also be expected for sMLE and MLE. Nonlinear dynamics arising from global bifurcations cannot be captured with a linear analysis near fixed points, and one would need to resort to numerical simulations; we explore these in section~\ref{sec:res}. To investigate into these nonlinear behaviors, we would need to numerically solve the three models. It is straightforward to solve the ODEs corresponding to LE and sMLE and we do this in MATLAB using the inbuilt solver ode45. However numerically solving the integro-differential equations for MLE is not trivial. We describe this numerical procedure in the next section.

\section{Numerical Integration of MLE}\label{sec:num}
In order to compute the nonlinear dynamics of the MLE (\ref{xeq})-(\ref{zeq}), we use a second order Adams-Bashforth predictor-corrector scheme for time-stepping, a second order method by Daitche \cite{Daitche2013} for the integrably singular part of $M(s)$, and a second order Newton-Cotes scheme for the remainder of $M(s)$. We will focus on (\ref{xeq}) as it introduces the most complexity for computation. We rewrite (\ref{xeq}) as
\begin{equation}
    \frac{\mathrm{d} X}{\mathrm{d} s} = Pr \left( Y - X - \frac{\mathrm{d}}{\mathrm{d}s}\int_0^s X(s')M(s - s') \mathrm{d} s' \right),
    \label{eqn:numericsfull}
\end{equation}
which is valid if $X(0) = 0$, which physically corresponds to the particle being stationary initially. Next, we split (\ref{eqn:numericsfull}) into three parts,
\begin{eqnarray}
    A &=& Pr\left(Y - X\right), \\
    B_1 &=& -\frac{\gamma Pr}{3 \sqrt{\pi}} \frac{\mathrm{d}}{\mathrm{d} s} \int_0^s \frac{X(s')}{\sqrt{s - s'}} \mathrm{d} s', \\
    B_2 &=& \frac{Pr}{3}\frac{\mathrm{d}}{\mathrm{d} s} \int_0^s X(s') \mathrm{erfc}\left( \sqrt{\frac{s - s'}{\gamma}}\right) \mathrm{exp} \left( \frac{s-s'}{\gamma} \right) \mathrm{d}s',
\end{eqnarray}
where $B_1$ contains the integrably singular part of $M(s)$. The time stepping scheme for $X(s)$ at time $s_{n+1} = s_n + h$, where $h$ is the step size, is
\begin{eqnarray}
    X(s_{n+1}) &=& X(s_n) + \frac{h}{2} \left(3 A(s_n) - A(s_{n-1}) \right) \nonumber \\
    &-& \zeta_1 \sum_{j = 0}^n \left(\beta_{j+1}^{n+1}X(s_{n-j}) - \beta_{j}^n X(s_{n-j}) \right) \nonumber \\
    &+& \zeta_2 \sum_{j = 0}^n \left(\eta_{j+1}^{n+1}X(s_{n-j}) - \eta_{j}^n X(s_{n-j}) \right),
    \label{eqn:Xnumerics}
\end{eqnarray}
where the second order coefficients $\beta_j^n$ are defined on page 6 of \cite{Daitche2013}, second order coefficients $\eta_j^n$ are the second order Newton Cotes coefficients, $\zeta_1 = \gamma Pr \sqrt{h}/(3 \sqrt{\pi})$, and $\zeta_2 = Pr/3$. The second term in (\ref{eqn:Xnumerics}) pertains to the integration of $B_1$ and the third term in (\ref{eqn:Xnumerics}) pertains to the integration of $B_2$. The time steppings for (\ref{yeq}) and (\ref{zeq}) are
\begin{eqnarray}
    Y(s_{n+1}) &=& Y(s_n) + \frac{h}{2} \Big[3\left(-X(s_n)Z(s_n) + r X(s_n) - Y(s_n) \right) \nonumber \\ 
    &-& \left(-X(s_{n-1})Z(s_{n-1}) + r X(s_{n-1}) - Y(s_{n-1}) \right) \Big],
\end{eqnarray}
\begin{eqnarray}
    Z(s_{n+1}) &=& Z(s_n) + \frac{h}{2} \Big[3\left(X(s_n)Y(s_n) - Z(s_n) \right) \nonumber \\ 
    &-& \left(X(s_{n-1})Y(s_{n-1}) - Z(s_{n-1}) \right) \Big].
\end{eqnarray}

\section{Nonlinear Quincke Rotor Dynamics}\label{sec:res}

We now turn to explore the nonlinear dynamics of the Quincke rotor as described by MLE in \eqref{xeq}-\eqref{zeq}, and also compare it with the resulting dynamics from the sMLE system in \eqref{eq: x sMLE}-\eqref{eq: z sMLE} and the standard Lorenz equations (LE) (i.e.\eqref{xeq}-\eqref{zeq} without the integral term in \eqref{xeq}). 

\begin{figure}
  \centerline{\includegraphics[width = \columnwidth]{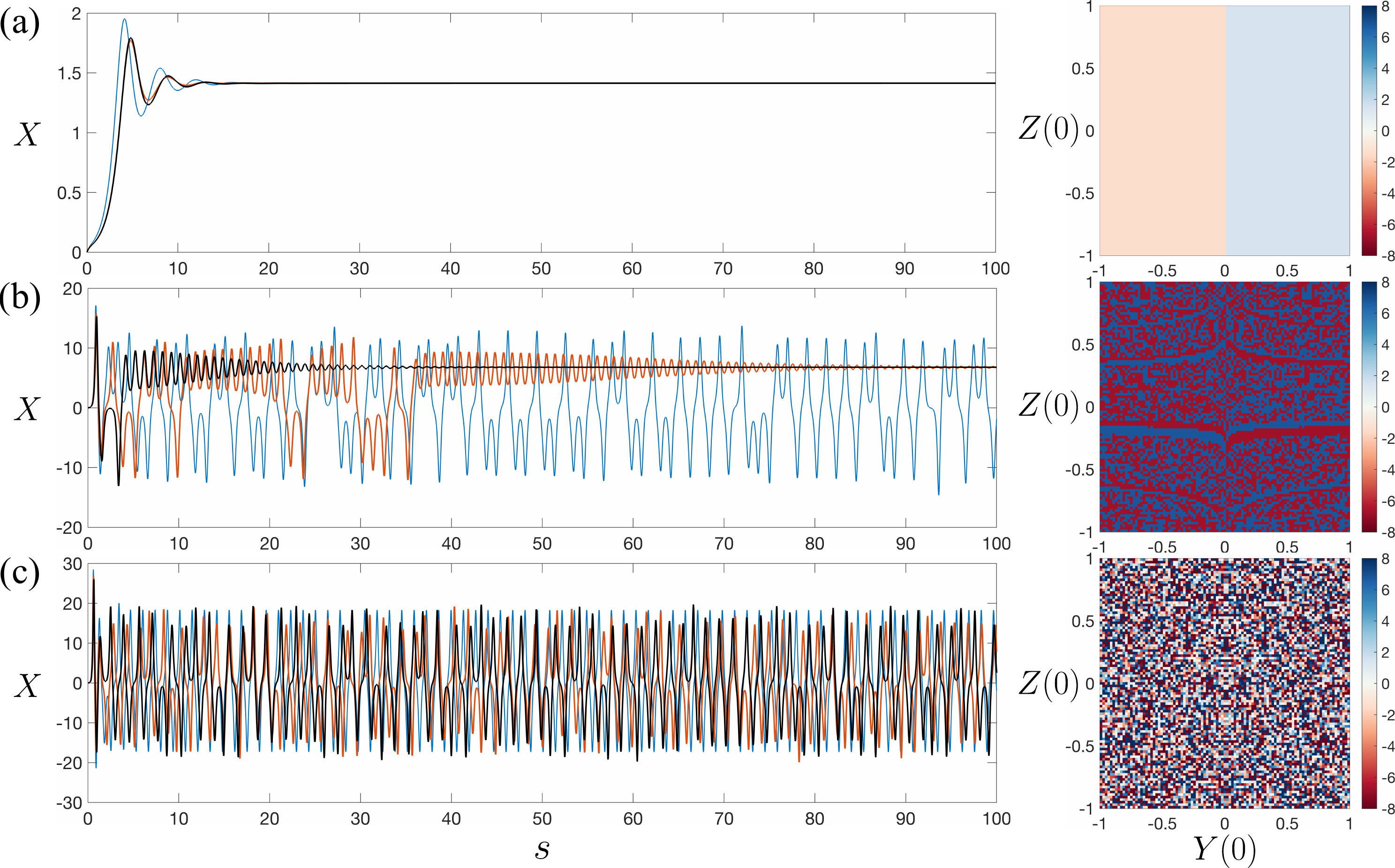}}
  \caption{(Left) Time series of the dimensionless angular velocity $X$ of the Quincke rotor for (blue) standard Lorenz system, (red) sMLE with $\alpha=0.5$, and (black) MLE with $\gamma=1$ at the dimensionless applied electric field strength (a) $r=3$, (b) $r=47$ and (c) $r=110$. (Right) Corresponding basin of attraction plots in the initial condition space formed by $(Y(0),Z(0))$ for sMLE model showing the final value of $X$ at the end of the simulation (with simulation time of $s=500$) for fixed $X(0)=0$. We see a smoothly separated basin for $r=3$, fractal basins due to transient chaos for $r=47$, and noise in the plot for $r=110$ due to permanent chaos. The parameter $Pr$ was fixed to $2.5$. Supplemental Videos 1-3 show the dynamics of MLE at $r=3, 47$ and $110$, respectively.}
\label{fig:TimeSeries}
\end{figure}

We first calculate the time-series dynamics of a spherical particle undergoing Quincke rotation with $\gamma = 1$ in MLE for three values of $r$, the dimensionless applied electric field strength. We compare these results at the same $r$ value for sMLE (with $\alpha=0.5$) and LE models. In figure \ref{fig:TimeSeries}, the time series for all three models is shown for $r = 3$, $47$, and $110$, demonstrating steady rotation ($r = 3$); transient chaos (or `preturbulence') before steady rotation ($r = 47$); and permanent chaotic dynamics ($r = 110)$. For $r=3$, we find that all three models, LE (blue), sMLE (red) and MLE (black), predict a steady rotation~(see figure \ref{fig:TimeSeries}(a)). This steady state rotation value is given by $X=\pm\sqrt{r-1}$ from the steady equilibrium states of the three models. At this $r$ value, a basin of attraction plotting the final value of $X$ in the initial condition space formed by $Y(0)$ and $Z(0)$ using sMLE equations is shown in the right panel of figure \ref{fig:TimeSeries}(a). We see that the basin of clockwise and counterclockwise steady rotating states are smoothly partitioned.

For a larger value of $r=47$, as shown in figure \ref{fig:TimeSeries}(b), we find that sMLE and MLE settles eventually on the steady value of $X$ after a chaotic transient, whereas LE, which does not include hydrodynamic memory, already exhibits permanent chaotic motion. Thus, we already see a signature of delay in the onset of chaotic motion with the inclusion of a memory term in the Lorenz equations. For LE at $Pr=2.5$, the steady rotation branch becomes unstable at $r = 32.5$; hence, at $r=47$ the only attracting set in phase space is a strange attractor resulting in chaotic dynamics. We note that the presence of transient chaos for sMLE and MLE indicates a final state sensitivity based on the initial conditions. Unlike $r=3$, where a $Y(0)>0$ would always result in counterclockwise steady rotation and $Y(0)<0$ in clockwise steady rotation, now at $r=47$, whether a clockwise or a counterclockwise steady rotation will be achieved in sMLE and MLE depends sensitively on initial conditions due to transient chaos. This can be seen in the basin of attraction plot of sMLE in the right panel of figure \ref{fig:TimeSeries}(b), where the basins of the two steady states are intricately mixed with fractal boundaries; a characteristic feature of transient chaos~\cite{RevModPhys.81.333}. 

For an even larger value of $r=110$, as shown in figure \ref{fig:TimeSeries}(c), the sMLE and MLE systems have now transitioned to permanent chaos, whereas the LE shows a periodic trajectory. The chaotic nature of the trajectory results in a noisy basin plot for sMLE as shown in the right panel of figure \ref{fig:TimeSeries}(c), where the final velocity at the end of the simulation can now take any value and is not correlated with the initial conditions.


\begin{figure}
  \centerline{\includegraphics[width = 0.5\columnwidth]{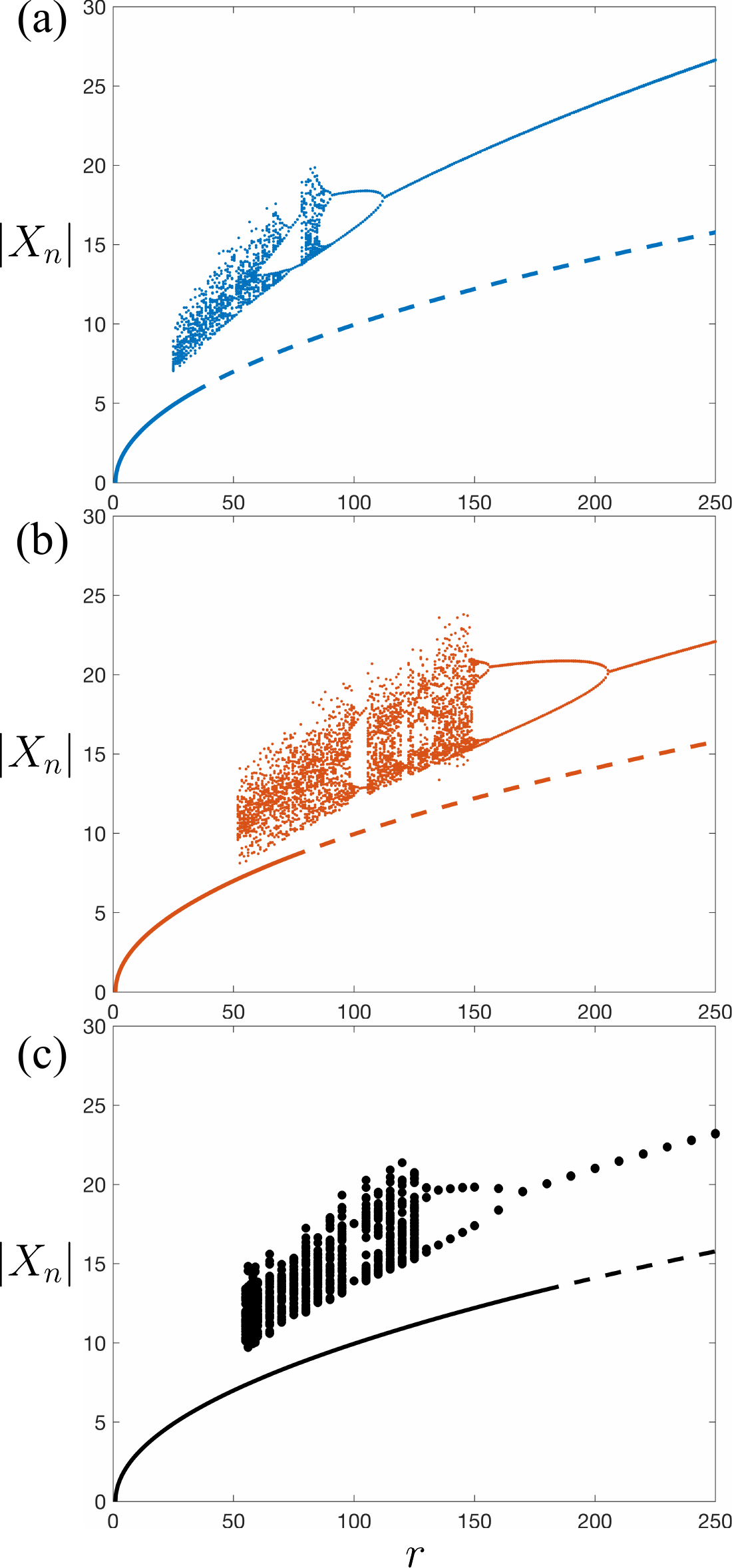}}
  \caption{Bifurcation diagram showing the peaks of $|X|$ i.e. $|X_n|$ as a function of $r$ and fixed $Pr=2.5$ for (a) standard LE, (b) sMLE with $M(s)=\alpha\, \text{e}^{-s/\alpha}$ and $\alpha=0.5$, and (c) MLE with $\gamma=1$. For each plot, the lower branch is the steady rotation state with the solid curve indicating where it is stable and the dashed curve indicating its unstable. The steady branch becomes unstable at $r=32.5$, $r\approx 73.4$ and $r\approx 176$ for LE, sMLE and MLE, respectively. Note the multistability regions where both chaotic motion and steady rotation coexist. The onset of multistability occurs at $r\approx 25$, $r\approx 51$ and $r\approx 55$ for LE, sMLE and MLE, respectively.}
\label{fig:Bifurcation}
\end{figure}

In figure \ref{fig:Bifurcation} we present a bifurcation diagram showing the peaks of $|X|$ as a function of $r$ and fixed $Pr=2.5$. Here the numerical integration was conducted until a time $s=500$ for the LE and sMLE, and $s=400$ for the MLE (due to the lengthier computations associated with numerical integration of the MLE). For the LE the steady rotation branch becomes unstable at $r= 32.5$, but a multistability region where both chaotic motion and steady rotation coexist occurs below this value of $r$. Indeed, chaotic dynamics are first observed at $r\approx 25$. The onset of chaotic dynamics (coexisting with steady rotation) is seen at $r\approx 55$ for the MLE (with $\gamma=1)$, which is a significantly larger $r$ value than for the LE. This larger $r$ value for the second bifurcation in the MLE is also seen in experiments by Peters \cite{Peters2005}; that is, the onset of chaos in their experiments occurs at a larger $r$ value than that predicted by the LE, and the present work suggests that hydrodynamic memory is the cause. The sMLE also have an increased $r$ value for the onset of chaotic dynamics (coexisiting with steady rotation) of $r\approx 51$, which suggests this might be a generic consequence of hydrodynamic memory, irrespective of the precise form of the memory kernel.

\begin{figure}
  \centerline{\includegraphics[width = \columnwidth]{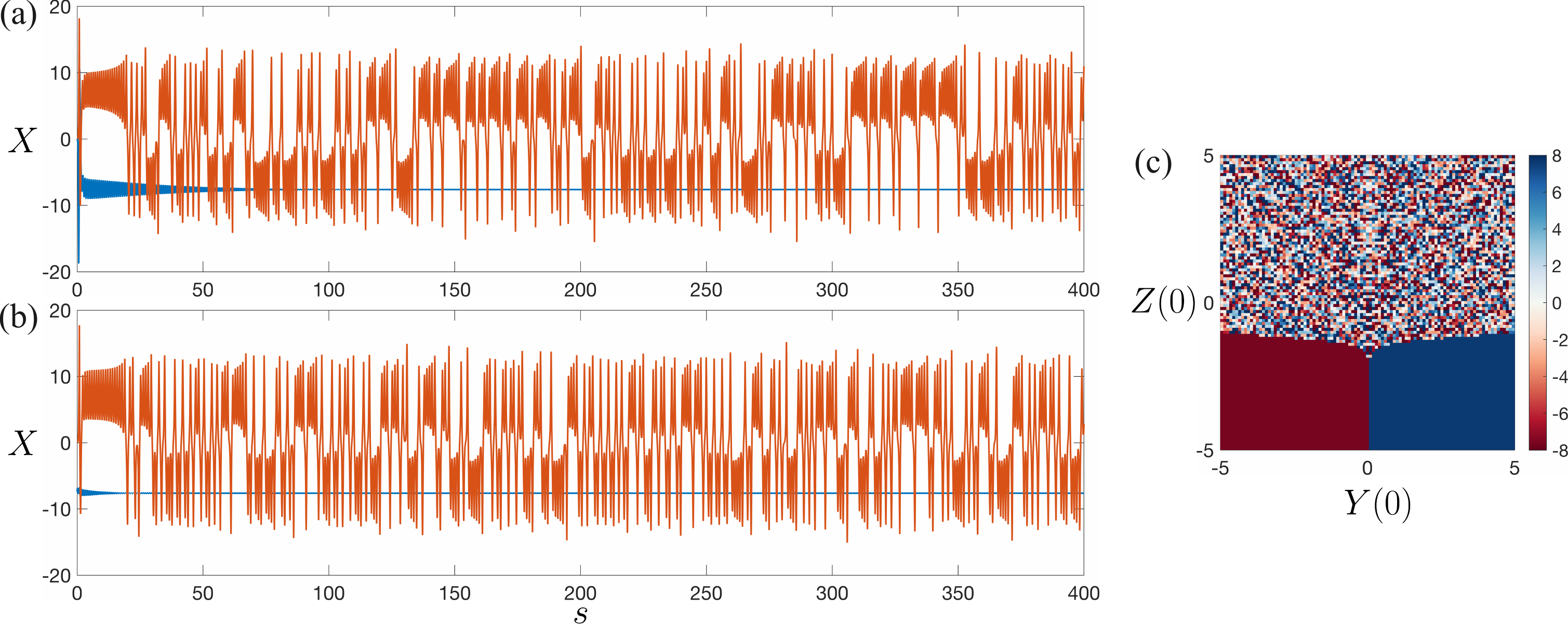}}
  \caption{Multistability at $Pr=2.5$ and $r=59$ in the (a) sMLE model with $\alpha=0.5$ and (b) MLE with $\gamma=1$. Time series of $X$ for two different initial conditions where one asymptotes towards a chaotic attractor (red) and the other towards a steady state (blue). (c) Corresponding basin of attraction plot in the initial condition space formed by $(Y(0),Z(0))$ for sMLE model showing the final value of $X$ for fixed $X(0)=0$. Supplemental Videos 4 and 5 show for MLE model the corresponding chaotic motion and steady dynamics, respectively.}
\label{fig:Multistability SMLE}
\end{figure}

In figure \ref{fig:Multistability SMLE} the multistability of the sMLE (with $\alpha=0.5$, figure \ref{fig:Multistability SMLE}(a)) and MLE (with $\gamma = 1$, figure \ref{fig:Multistability SMLE}(b)) are examined by plotting the time series of $X$ for two different initial conditions, where one asymptotes towards a chaotic attractor (red) and the other towards a steady state (blue). 
Figure \ref{fig:Multistability SMLE}(c) displays basin of attraction plot in the initial condition space $(Y(0),Z(0))$ in the sMLE model for fixed $X(0)=0$ showing the final value of $X$ at the end of simulation time of $400$. We can see coexistence of steady states with smooth separation of their basins, which reside alongside chaotic solutions with noisy values in the basin. Thus, in the multistable region, we see that understanding the size and organization of basin of attraction of steady rotation and chaotic motion might assist with predicting which of the two final states is likely to be realized.

\begin{figure}
  \centerline{\includegraphics[width = 0.8\columnwidth]{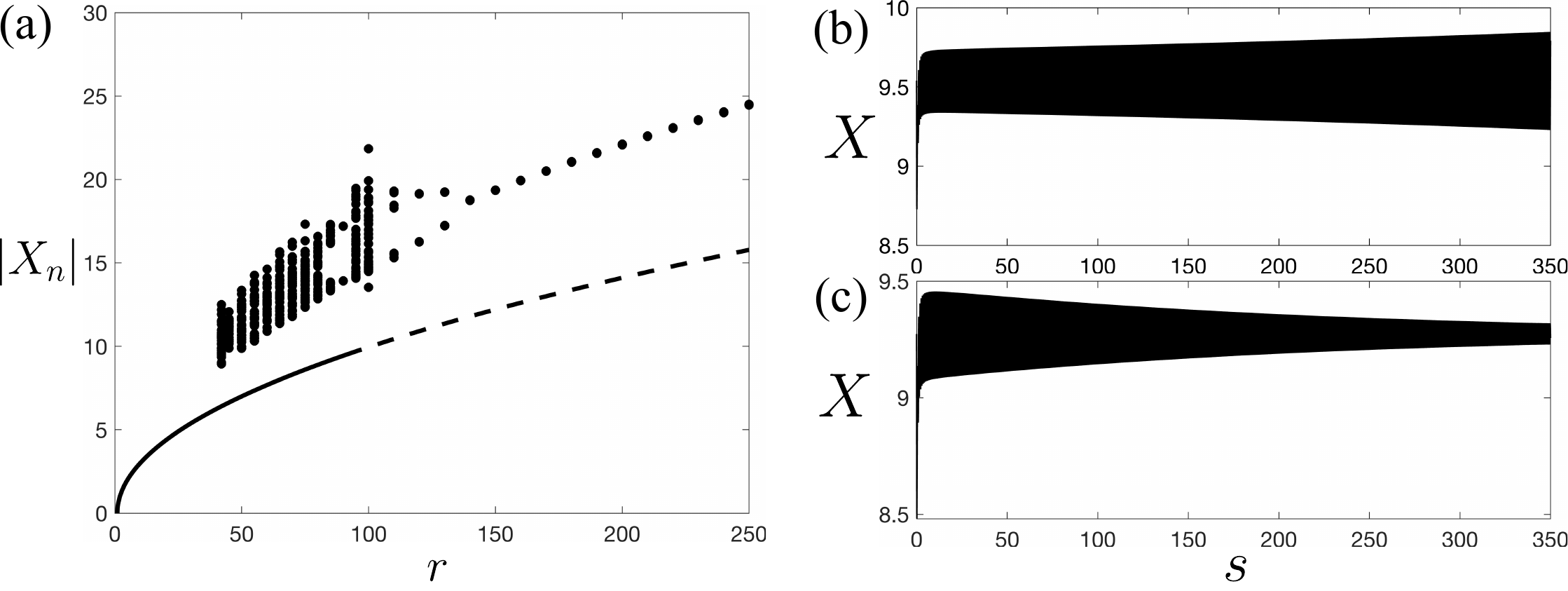}}
  \caption{(a) Bifurcation diagram showing the peaks of $|X|$ i.e. $|X_n|$ as a function of $r$ and fixed $Pr=2.5$ for full MLE with $\gamma=0.5$. The lower branch is the steady rotation state with the solid curve indicating where it is stable and the dashed curve indicating its unstable. (b) Time series of $X$ just (b) before ($r=87$) and (c) after ($r=92$) the end of the stability of the steady rotation branch.}
\label{fig:Bifurcation 2}
\end{figure}

In figure \ref{fig:Bifurcation 2}(a) we present a bifurcation diagram of the MLE showing the peaks of $|X|$ as a function of $r$ at fixed $Pr=2.5$ and $\gamma =0.5$. Here, the onset of chaos is at $r\approx 42$ and there is again a region of multistability in which chaotic and steady motion coexist (with steady rotation stable up to $r\approx 90$). In figure \ref{fig:Bifurcation 2}(b) and figure \ref{fig:Bifurcation 2}(c) we plot the times series of $X$ just (b) before ($r=87$) and (c) after ($r=92$), respectively, the end of the stability of the steady rotation branch. At $r=87$ there is a slow convergence of $X$ toward its steady value, and for $r=92$ there is a slow growth of $X$, which we envision would at some time point beyond the simulation become chaotic. The value of $\gamma =0.5$ was chosen as this is approximately the same value as in the experiments of \cite{Peters2005}. They observed chaotic dynamics at a field strength of $6.5\mathrm{kV/cm}$, whereas the LE would predict chaos to onset at $5.5\mathrm{kV/cm}$, which corresponds to the ratio of $r$ values (experiment to LE) of around $1.4$. Our computations at $\gamma =0.5$ give a ratio of $r$ values (MLE to LE) of $42/32.5\approx 1.3$ for the onset of chaos. This corresponds to a dimensional onset field strength of $6.3\mathrm{kV/cm}$ for the MLE. Thus the ratio of $r$ values (experiment to MLE) is around $1.06$, which is in encouraging agreement.

\begin{figure}
  \centerline{\includegraphics[width = \columnwidth]{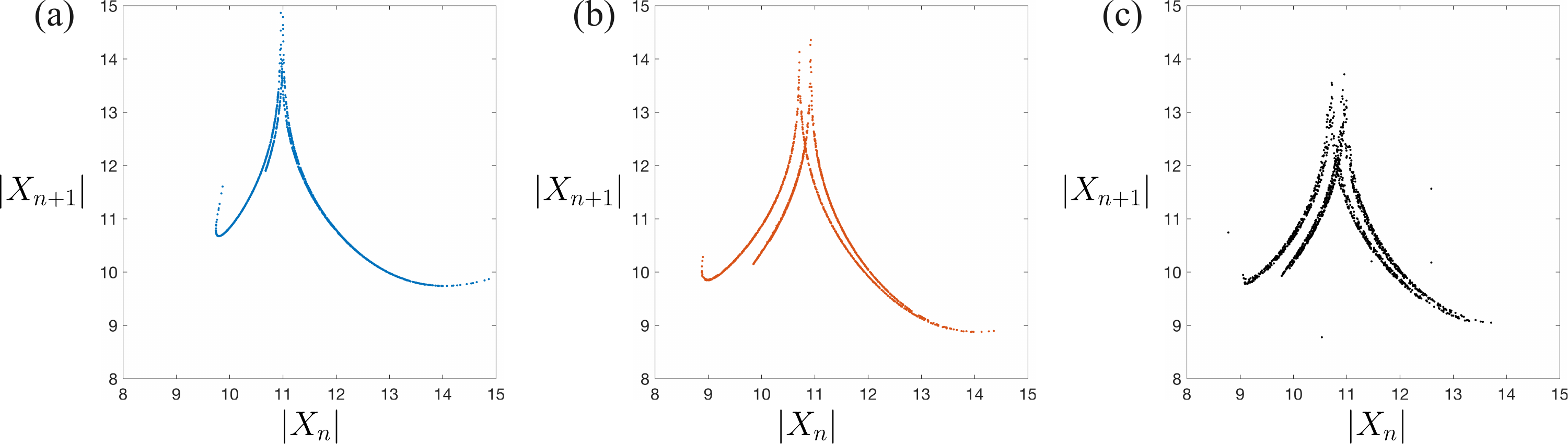}}
  \caption{First return map for the peaks in time series of the magnitude of $X$ for $r=45$ and fixed $Pr=2.5$ for (a) LE, (b) sMLE with $\alpha=0.25$, and (c) full MLE with $\gamma=0.5$.}
\label{fig:returnmap}
\end{figure}

To further compare the nature of chaotic dynamics near the onset of chaos between the three models, we plot the first-return map of peaks of $|X|$ in Fig.~\ref{fig:returnmap} for LE, sMLE and MLE models at $r=45$ and $Pr=2.5$. As it can be seen in Fig.~\ref{fig:returnmap}(a), the LE model shows the characteristic cusp-like structure of the first return map that is commonly reported for Lorenz chaos. Alternatively, as shown in Figs.~\ref{fig:returnmap}(b)-(c), for sMLE and MLE respectively, we find a double-cusp structure in the first return map. This suggests that the double-cusp structure might be a consequence of memory effects in our extended Lorenz sMLE and MLE models. Furthermore, it is not entirely clear whether the cusp structure in the experimental first return map of \cite{Peters2005} (see their Fig.~7) has a single cusp or a closely spaced double-cusp structure. This feature might be a useful way to characterize memory effects in chaotic Quincke rotation experiments.

\begin{figure}
  \centerline{\includegraphics[width = \columnwidth]{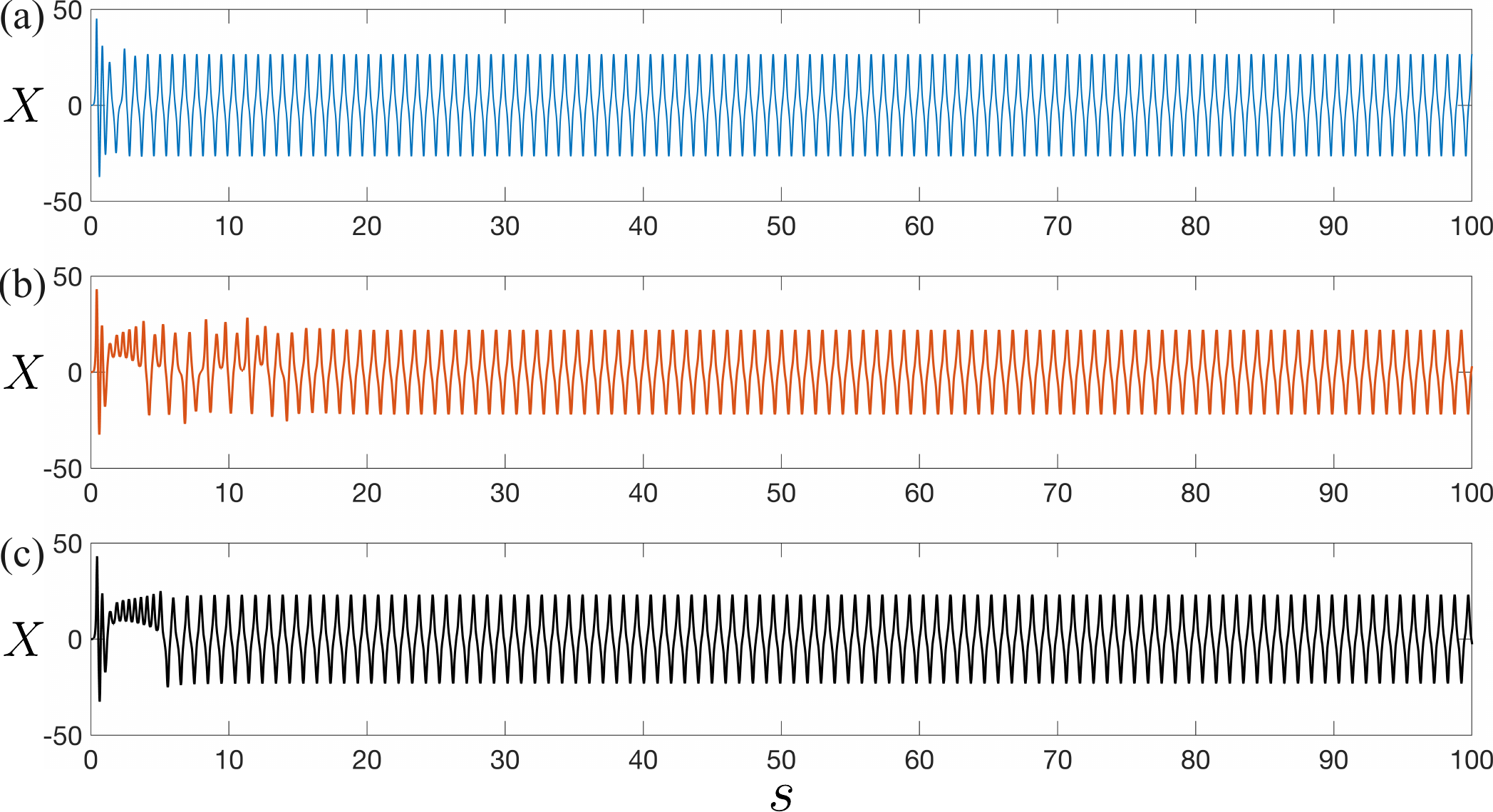}}
  \caption{Time series of $X$ showing periodic oscillations for large $r=250$ and fixed $Pr=2.5$ for (a) LE, (b) sMLE with $\alpha=0.5$, and (c) full MLE with $\gamma=1$. Supplemental Video 6 shows the dynamics of MLE in this periodic state.}
\label{fig:periodicorbit}
\end{figure}

In figure \ref{fig:periodicorbit} the time series of $X$ is plotted at a large value of $r=250$ for the LE, sMLE with $\alpha=0.5$, and MLE with $\gamma =1$. All three series converge onto a periodic orbit; however, the orbits are not in phase. It is well known for LE that periodic solutions exist at large $r$ and the system becomes integrable in this limit~\citep{jackson_1990}. It appears that we get the same limiting behavior for sMLE and MLE for large $r$. Specifically, we find that for all three models, the period of oscillations decrease whereas the amplitude of oscillations increase, with increasing $r$. Furthermore, we find that the period of the Lorenz-type systems with memory (sMLE and MLE) is larger than that of LE, and the amplitude of the Lorenz-type systems with memory is smaller than that of LE, at a given $r$ in this periodic regime. It would be interesting to explore this regime experimentally and verify if periodic rotations are obtained for Quincke rotors under a large magnitude of applied electric field strengths.

\section{Conclusion}\label{sec:con}

A mathematical model for the Quincke rotation of a spherical particle that accounts for the inertia of the particle and the unsteady Stokes flow in the surrounding fluid is presented and analyzed.
This model takes the form of an integro-differential, infinite-dimensional dynamical system that we term the `memory Lorenz equations' (MLE).
The primary conclusion is that the linearized fluid inertia, or hydrodynamic memory as we call it, causes an increase in the threshold field strength for observation of chaotic rotor dynamics.
This is in qualitative agreement with previous experimental observations.
A secondary conclusion is that increasing the ratio of momentum diffusion to dipole relaxation times ($\gamma$) leads to an increasing range of field strengths for which multi-stability, or co-existence, between steady and chaotic rotation occurs.
In the experiments of \cite{Peters2005}, the radius of the cylindrical rotor (a glass tube) is around 1mm, whereas the spacing between the electrodes that supply the electric field is around 1cm. The rotor is placed at the midpoint of the electrodes. Hence, a more accurate theoretical treatment would require accounting for the hydrodynamic interactions between the rotor and the electrodes, which would alter the form of the memory kernel in the MLE.
However, we do not expect the primary and secondary conclusions to be affected by our neglect of such interactions.
Further, spherical particles often undergo Quincke rotation near a wall (upon which they have settled), in which case they are referred to as `Quincke rollers,' since the rotation is accompanied by translational motion along the wall. Certainly, the hydrodynamic interaction between the particle and wall would affect the hydrodynamic memory in this case.

A simplified form of the MLE (i.e., sMLE), which may be relevant to rotor dynamics in non-Newtonian fluids, indicates that the above conclusions could be a generic feature of Lorenz-type systems with distributed memory. {
This method of suppressing the onset of chaos with memory effects may be relevant to the field of chaos control~\citep{Pyragas2001PRL,Pyragas2006PRE} where time-delayed feedback is routinely used to control chaotic dynamics in various systems. Further, since the effects of exponentially fading memory for sMLE are captured by just one additional ODE coupled to the original $3$D Lorenz equations, this dynamical system might provide useful insights into suitable chaos control methods using simple feedback mechanisms~\citep{pyragas2006royal}.}

It would, of course, be desirable to test in more detail our predictions against fresh experiments; for example, can the co-existence between steady and chaotic rotation be observed experimentally?
Since the MLE model appears to preserve all the intricate dynamical features of LE, experiments with a Quincke rotor might form a testing bed to experimentally realize the complex bifurcations associated with Lorenz chaos. 
Natural extensions of the present work include rotor dynamics in time-dependent fields, or dynamics in the presence of a background shear flow.


\section*{Acknowledgments}
R.V. acknowledges the support of the Leverhulme Trust [Grant No. LIP-2020-014] and the ERC Advanced Grant ActBio (funded as UKRI Frontier Research Grant EP/Y033981/1).
J. K. acknowledges the support Sharbaugh Presidential Fellowship and Dighe Fellowship from the Department of Chemical Engineering at Carnegie Mellon University.

\bibliography{7refs}

\providecommand{\noopsort}[1]{}\providecommand{\singleletter}[1]{#1}%
\begin{thebibliography}{35}%
\makeatletter
\providecommand \@ifxundefined [1]{%
 \@ifx{#1\undefined}
}%
\providecommand \@ifnum [1]{%
 \ifnum #1\expandafter \@firstoftwo
 \else \expandafter \@secondoftwo
 \fi
}%
\providecommand \@ifx [1]{%
 \ifx #1\expandafter \@firstoftwo
 \else \expandafter \@secondoftwo
 \fi
}%
\providecommand \natexlab [1]{#1}%
\providecommand \enquote  [1]{``#1''}%
\providecommand \bibnamefont  [1]{#1}%
\providecommand \bibfnamefont [1]{#1}%
\providecommand \citenamefont [1]{#1}%
\providecommand \href@noop [0]{\@secondoftwo}%
\providecommand \href [0]{\begingroup \@sanitize@url \@href}%
\providecommand \@href[1]{\@@startlink{#1}\@@href}%
\providecommand \@@href[1]{\endgroup#1\@@endlink}%
\providecommand \@sanitize@url [0]{\catcode `\\12\catcode `\$12\catcode `\&12\catcode `\#12\catcode `\^12\catcode `\_12\catcode `\%12\relax}%
\providecommand \@@startlink[1]{}%
\providecommand \@@endlink[0]{}%
\providecommand \url  [0]{\begingroup\@sanitize@url \@url }%
\providecommand \@url [1]{\endgroup\@href {#1}{\urlprefix }}%
\providecommand \urlprefix  [0]{URL }%
\providecommand \Eprint [0]{\href }%
\providecommand \doibase [0]{https://doi.org/}%
\providecommand \selectlanguage [0]{\@gobble}%
\providecommand \bibinfo  [0]{\@secondoftwo}%
\providecommand \bibfield  [0]{\@secondoftwo}%
\providecommand \translation [1]{[#1]}%
\providecommand \BibitemOpen [0]{}%
\providecommand \bibitemStop [0]{}%
\providecommand \bibitemNoStop [0]{.\EOS\space}%
\providecommand \EOS [0]{\spacefactor3000\relax}%
\providecommand \BibitemShut  [1]{\csname bibitem#1\endcsname}%
\let\auto@bib@innerbib\@empty
\bibitem [{\citenamefont {Quincke}(1896)}]{Quincke1896}%
  \BibitemOpen
  \bibfield  {author} {\bibinfo {author} {\bibfnamefont {G.}~\bibnamefont {Quincke}},\ }\bibfield  {title} {\bibinfo {title} {Ueber rotationen im constanten electrischen felde},\ }\href@noop {} {\bibfield  {journal} {\bibinfo  {journal} {Annalen der Physik}\ }\textbf {\bibinfo {volume} {295}},\ \bibinfo {pages} {417} (\bibinfo {year} {1896})}\BibitemShut {NoStop}%
\bibitem [{\citenamefont {Dolinsky~Y}(2012)}]{Dolinsky2012}%
  \BibitemOpen
  \bibfield  {author} {\bibinfo {author} {\bibfnamefont {E.~T.}\ \bibnamefont {Dolinsky~Y}},\ }\bibfield  {title} {\bibinfo {title} {Dipole interaction of the quincke rotating particles},\ }\href@noop {} {\bibfield  {journal} {\bibinfo  {journal} {Phys Rev E Stat Nonlin Soft Matter Phys.}\ }\textbf {\bibinfo {volume} {85}},\ \bibinfo {pages} {026608} (\bibinfo {year} {2012})}\BibitemShut {NoStop}%
\bibitem [{\citenamefont {Sherman}\ and\ \citenamefont {Swan}(2020)}]{Sherman2020}%
  \BibitemOpen
  \bibfield  {author} {\bibinfo {author} {\bibfnamefont {Z.~M.}\ \bibnamefont {Sherman}}\ and\ \bibinfo {author} {\bibfnamefont {J.~W.}\ \bibnamefont {Swan}},\ }\bibfield  {title} {\bibinfo {title} {Spontaneous electrokinetic magnus effect},\ }\href@noop {} {\bibfield  {journal} {\bibinfo  {journal} {Phys. Rev. Lett.}\ }\textbf {\bibinfo {volume} {124}},\ \bibinfo {pages} {208002} (\bibinfo {year} {2020})}\BibitemShut {NoStop}%
\bibitem [{\citenamefont {Salipante}\ and\ \citenamefont {Vlahovska}(2010)}]{Salipante2010}%
  \BibitemOpen
  \bibfield  {author} {\bibinfo {author} {\bibfnamefont {P.~F.}\ \bibnamefont {Salipante}}\ and\ \bibinfo {author} {\bibfnamefont {P.~M.}\ \bibnamefont {Vlahovska}},\ }\bibfield  {title} {\bibinfo {title} {Electrohydrodynamics of drops in strong uniform dc electric fields},\ }\href@noop {} {\bibfield  {journal} {\bibinfo  {journal} {Physics of Fluids}\ }\textbf {\bibinfo {volume} {22}},\ \bibinfo {pages} {112110} (\bibinfo {year} {2010})}\BibitemShut {NoStop}%
\bibitem [{\citenamefont {Ouriemi}\ and\ \citenamefont {Vlahovska}(2015)}]{Ouriemi2015}%
  \BibitemOpen
  \bibfield  {author} {\bibinfo {author} {\bibfnamefont {M.}~\bibnamefont {Ouriemi}}\ and\ \bibinfo {author} {\bibfnamefont {P.~M.}\ \bibnamefont {Vlahovska}},\ }\bibfield  {title} {\bibinfo {title} {Electrohydrodynamic deformation and rotation of a particle-coated drop},\ }\href@noop {} {\bibfield  {journal} {\bibinfo  {journal} {Langmuir}\ }\textbf {\bibinfo {volume} {31}},\ \bibinfo {pages} {6298} (\bibinfo {year} {2015})}\BibitemShut {NoStop}%
\bibitem [{\citenamefont {Vlahovska}(2019)}]{Vlahovska2019}%
  \BibitemOpen
  \bibfield  {author} {\bibinfo {author} {\bibfnamefont {P.~M.}\ \bibnamefont {Vlahovska}},\ }\bibfield  {title} {\bibinfo {title} {Electrohydrodynamics of drops and vesicles},\ }\href@noop {} {\bibfield  {journal} {\bibinfo  {journal} {Annual Review of Fluid Mechanics}\ }\textbf {\bibinfo {volume} {51}},\ \bibinfo {pages} {305} (\bibinfo {year} {2019})}\BibitemShut {NoStop}%
\bibitem [{\citenamefont {Bricard}\ \emph {et~al.}(2015)\citenamefont {Bricard}, \citenamefont {Caussin}, \citenamefont {Das}, \citenamefont {Savoie}, \citenamefont {Chikkadi}, \citenamefont {Shitara}, \citenamefont {Chepizhko}, \citenamefont {Peruani}, \citenamefont {Saintillan},\ and\ \citenamefont {Bartolo}}]{Bricard2015}%
  \BibitemOpen
  \bibfield  {author} {\bibinfo {author} {\bibfnamefont {A.}~\bibnamefont {Bricard}}, \bibinfo {author} {\bibfnamefont {J.-B.}\ \bibnamefont {Caussin}}, \bibinfo {author} {\bibfnamefont {D.}~\bibnamefont {Das}}, \bibinfo {author} {\bibfnamefont {C.}~\bibnamefont {Savoie}}, \bibinfo {author} {\bibfnamefont {V.}~\bibnamefont {Chikkadi}}, \bibinfo {author} {\bibfnamefont {K.}~\bibnamefont {Shitara}}, \bibinfo {author} {\bibfnamefont {O.}~\bibnamefont {Chepizhko}}, \bibinfo {author} {\bibfnamefont {F.}~\bibnamefont {Peruani}}, \bibinfo {author} {\bibfnamefont {D.}~\bibnamefont {Saintillan}},\ and\ \bibinfo {author} {\bibfnamefont {D.}~\bibnamefont {Bartolo}},\ }\bibfield  {title} {\bibinfo {title} {Emergent vortices in populations of colloidal rollers},\ }\href@noop {} {\bibfield  {journal} {\bibinfo  {journal} {Nature Communications}\ }\textbf {\bibinfo {volume} {6}},\ \bibinfo {pages} {7470} (\bibinfo {year} {2015})}\BibitemShut {NoStop}%
\bibitem [{\citenamefont {Das}\ and\ \citenamefont {Lauga}(2019)}]{Das2019}%
  \BibitemOpen
  \bibfield  {author} {\bibinfo {author} {\bibfnamefont {D.}~\bibnamefont {Das}}\ and\ \bibinfo {author} {\bibfnamefont {E.}~\bibnamefont {Lauga}},\ }\bibfield  {title} {\bibinfo {title} {Active particles powered by quincke rotation in a bulk fluid},\ }\href {https://doi.org/10.1103/PhysRevLett.122.194503} {\bibfield  {journal} {\bibinfo  {journal} {Phys. Rev. Lett.}\ }\textbf {\bibinfo {volume} {122}},\ \bibinfo {pages} {194503} (\bibinfo {year} {2019})}\BibitemShut {NoStop}%
\bibitem [{\citenamefont {Pradillo}\ \emph {et~al.}(2019)\citenamefont {Pradillo}, \citenamefont {Karani},\ and\ \citenamefont {Vlahovska}}]{Pradillo2019}%
  \BibitemOpen
  \bibfield  {author} {\bibinfo {author} {\bibfnamefont {G.~E.}\ \bibnamefont {Pradillo}}, \bibinfo {author} {\bibfnamefont {H.}~\bibnamefont {Karani}},\ and\ \bibinfo {author} {\bibfnamefont {P.~M.}\ \bibnamefont {Vlahovska}},\ }\bibfield  {title} {\bibinfo {title} {Quincke rotor dynamics in confinement: rolling and hovering},\ }\href@noop {} {\bibfield  {journal} {\bibinfo  {journal} {Soft Matter}\ }\textbf {\bibinfo {volume} {15}},\ \bibinfo {pages} {6564} (\bibinfo {year} {2019})}\BibitemShut {NoStop}%
\bibitem [{\citenamefont {Karani}\ \emph {et~al.}(2019)\citenamefont {Karani}, \citenamefont {Pradillo},\ and\ \citenamefont {Vlahovska}}]{Karani2019}%
  \BibitemOpen
  \bibfield  {author} {\bibinfo {author} {\bibfnamefont {H.}~\bibnamefont {Karani}}, \bibinfo {author} {\bibfnamefont {G.~E.}\ \bibnamefont {Pradillo}},\ and\ \bibinfo {author} {\bibfnamefont {P.~M.}\ \bibnamefont {Vlahovska}},\ }\bibfield  {title} {\bibinfo {title} {Tuning the random walk of active colloids: From individual run-and-tumble to dynamic clustering},\ }\href {https://doi.org/10.1103/PhysRevLett.123.208002} {\bibfield  {journal} {\bibinfo  {journal} {Phys. Rev. Lett.}\ }\textbf {\bibinfo {volume} {123}},\ \bibinfo {pages} {208002} (\bibinfo {year} {2019})}\BibitemShut {NoStop}%
\bibitem [{\citenamefont {Zhu}\ and\ \citenamefont {Stone}(2019)}]{Zhu2019}%
  \BibitemOpen
  \bibfield  {author} {\bibinfo {author} {\bibfnamefont {L.}~\bibnamefont {Zhu}}\ and\ \bibinfo {author} {\bibfnamefont {H.~A.}\ \bibnamefont {Stone}},\ }\bibfield  {title} {\bibinfo {title} {Propulsion driven by self-oscillation via an electrohydrodynamic instability},\ }\href {https://doi.org/10.1103/PhysRevFluids.4.061701} {\bibfield  {journal} {\bibinfo  {journal} {Phys. Rev. Fluids}\ }\textbf {\bibinfo {volume} {4}},\ \bibinfo {pages} {061701} (\bibinfo {year} {2019})}\BibitemShut {NoStop}%
\bibitem [{\citenamefont {C\ifmmode~\bar{e}\else \={e}\fi{}bers}\ \emph {et~al.}(2000)\citenamefont {C\ifmmode~\bar{e}\else \={e}\fi{}bers}, \citenamefont {Lemaire},\ and\ \citenamefont {Lobry}}]{Cebers2000}%
  \BibitemOpen
  \bibfield  {author} {\bibinfo {author} {\bibfnamefont {A.}~\bibnamefont {C\ifmmode~\bar{e}\else \={e}\fi{}bers}}, \bibinfo {author} {\bibfnamefont {E.}~\bibnamefont {Lemaire}},\ and\ \bibinfo {author} {\bibfnamefont {L.}~\bibnamefont {Lobry}},\ }\bibfield  {title} {\bibinfo {title} {Electrohydrodynamic instabilities and orientation of dielectric ellipsoids in low-conducting fluids},\ }\href {https://doi.org/10.1103/PhysRevE.63.016301} {\bibfield  {journal} {\bibinfo  {journal} {Phys. Rev. E}\ }\textbf {\bibinfo {volume} {63}},\ \bibinfo {pages} {016301} (\bibinfo {year} {2000})}\BibitemShut {NoStop}%
\bibitem [{\citenamefont {C\ifmmode~\bar{e}\else \={e}\fi{}bers}(2004)}]{Cebers2004}%
  \BibitemOpen
  \bibfield  {author} {\bibinfo {author} {\bibfnamefont {A.}~\bibnamefont {C\ifmmode~\bar{e}\else \={e}\fi{}bers}},\ }\bibfield  {title} {\bibinfo {title} {Bistability and ``negative'' viscosity for a suspension of insulating particles in an electric field},\ }\href {https://doi.org/10.1103/PhysRevLett.92.034501} {\bibfield  {journal} {\bibinfo  {journal} {Phys. Rev. Lett.}\ }\textbf {\bibinfo {volume} {92}},\ \bibinfo {pages} {034501} (\bibinfo {year} {2004})}\BibitemShut {NoStop}%
\bibitem [{\citenamefont {Lemaire}\ \emph {et~al.}(2008)\citenamefont {Lemaire}, \citenamefont {Lobry}, \citenamefont {Pannacci},\ and\ \citenamefont {Peters}}]{Lemaire2008}%
  \BibitemOpen
  \bibfield  {author} {\bibinfo {author} {\bibfnamefont {E.}~\bibnamefont {Lemaire}}, \bibinfo {author} {\bibfnamefont {L.}~\bibnamefont {Lobry}}, \bibinfo {author} {\bibfnamefont {N.}~\bibnamefont {Pannacci}},\ and\ \bibinfo {author} {\bibfnamefont {F.}~\bibnamefont {Peters}},\ }\bibfield  {title} {\bibinfo {title} {Viscosity of an electro-rheological suspension with internal rotations},\ }\href {https://doi.org/10.1122/1.2903546} {\bibfield  {journal} {\bibinfo  {journal} {Journal of Rheology}\ }\textbf {\bibinfo {volume} {52}} (\bibinfo {year} {2008})}\BibitemShut {NoStop}%
\bibitem [{\citenamefont {Belovs}\ and\ \citenamefont {C\ifmmode~\bar{e}\else \={e}\fi{}bers}(2020)}]{Belovs2020}%
  \BibitemOpen
  \bibfield  {author} {\bibinfo {author} {\bibfnamefont {M.}~\bibnamefont {Belovs}}\ and\ \bibinfo {author} {\bibfnamefont {A.}~\bibnamefont {C\ifmmode~\bar{e}\else \={e}\fi{}bers}},\ }\bibfield  {title} {\bibinfo {title} {Quincke rotation driven flows},\ }\href {https://doi.org/10.1103/PhysRevFluids.5.013701} {\bibfield  {journal} {\bibinfo  {journal} {Phys. Rev. Fluids}\ }\textbf {\bibinfo {volume} {5}},\ \bibinfo {pages} {013701} (\bibinfo {year} {2020})}\BibitemShut {NoStop}%
\bibitem [{\citenamefont {Pannacci}\ \emph {et~al.}(2007)\citenamefont {Pannacci}, \citenamefont {Lobry},\ and\ \citenamefont {Lemaire}}]{Pannacci2007}%
  \BibitemOpen
  \bibfield  {author} {\bibinfo {author} {\bibfnamefont {N.}~\bibnamefont {Pannacci}}, \bibinfo {author} {\bibfnamefont {L.}~\bibnamefont {Lobry}},\ and\ \bibinfo {author} {\bibfnamefont {E.}~\bibnamefont {Lemaire}},\ }\bibfield  {title} {\bibinfo {title} {How insulating particles increase the conductivity of a suspension},\ }\href {https://doi.org/10.1103/PhysRevLett.99.094503} {\bibfield  {journal} {\bibinfo  {journal} {Phys. Rev. Lett.}\ }\textbf {\bibinfo {volume} {99}},\ \bibinfo {pages} {094503} (\bibinfo {year} {2007})}\BibitemShut {NoStop}%
\bibitem [{\citenamefont {Das}\ and\ \citenamefont {Saintillan}(2013)}]{Das-13}%
  \BibitemOpen
  \bibfield  {author} {\bibinfo {author} {\bibfnamefont {D.}~\bibnamefont {Das}}\ and\ \bibinfo {author} {\bibfnamefont {D.}~\bibnamefont {Saintillan}},\ }\bibfield  {title} {\bibinfo {title} {{Electrohydrodynamic interaction of spherical particles under Quincke rotation}},\ }\href@noop {} {\bibfield  {journal} {\bibinfo  {journal} {Physical Review E}\ }\textbf {\bibinfo {volume} {87}},\ \bibinfo {pages} {194} (\bibinfo {year} {2013})}\BibitemShut {NoStop}%
\bibitem [{\citenamefont {Jones}(1984)}]{Jon-84}%
  \BibitemOpen
  \bibfield  {author} {\bibinfo {author} {\bibfnamefont {T.~B.}\ \bibnamefont {Jones}},\ }\bibfield  {title} {\bibinfo {title} {{Quincke Rotation of Spheres}},\ }\href@noop {} {\bibfield  {journal} {\bibinfo  {journal} {Industry Applications, IEEE Transactions on}\ ,\ \bibinfo {pages} {845}} (\bibinfo {year} {1984})}\BibitemShut {NoStop}%
\bibitem [{\citenamefont {Lemaire}\ and\ \citenamefont {Lobry}(2002)}]{Lemaire2002}%
  \BibitemOpen
  \bibfield  {author} {\bibinfo {author} {\bibfnamefont {E.}~\bibnamefont {Lemaire}}\ and\ \bibinfo {author} {\bibfnamefont {L.}~\bibnamefont {Lobry}},\ }\bibfield  {title} {\bibinfo {title} {Chaotic behavior in electro-rotation},\ }\href@noop {} {\bibfield  {journal} {\bibinfo  {journal} {Physica A: Statistical Mechanics and its Applications}\ }\textbf {\bibinfo {volume} {314}},\ \bibinfo {pages} {663} (\bibinfo {year} {2002})}\BibitemShut {NoStop}%
\bibitem [{\citenamefont {Peters}\ \emph {et~al.}(2005)\citenamefont {Peters}, \citenamefont {Lobry},\ and\ \citenamefont {Lemaire}}]{Peters2005}%
  \BibitemOpen
  \bibfield  {author} {\bibinfo {author} {\bibfnamefont {F.}~\bibnamefont {Peters}}, \bibinfo {author} {\bibfnamefont {L.}~\bibnamefont {Lobry}},\ and\ \bibinfo {author} {\bibfnamefont {E.}~\bibnamefont {Lemaire}},\ }\bibfield  {title} {\bibinfo {title} {Experimental observation of {L}orenz chaos in the {Q}uincke rotor dynamics},\ }\href@noop {} {\bibfield  {journal} {\bibinfo  {journal} {Chaos: An Interdisciplinary Journal of Nonlinear Science}\ }\textbf {\bibinfo {volume} {15}},\ \bibinfo {pages} {013102} (\bibinfo {year} {2005})}\BibitemShut {NoStop}%
\bibitem [{\citenamefont {Lorenz}(1963)}]{Lorenz1963}%
  \BibitemOpen
  \bibfield  {author} {\bibinfo {author} {\bibfnamefont {E.~N.}\ \bibnamefont {Lorenz}},\ }\bibfield  {title} {\bibinfo {title} {Deterministic nonperiodic flow},\ }\href@noop {} {\bibfield  {journal} {\bibinfo  {journal} {Journal of Atmospheric Sciences}\ }\textbf {\bibinfo {volume} {20}},\ \bibinfo {pages} {130 } (\bibinfo {year} {1963})}\BibitemShut {NoStop}%
\bibitem [{\citenamefont {Dommersnes}\ \emph {et~al.}(2016)\citenamefont {Dommersnes}, \citenamefont {Mikkelsen},\ and\ \citenamefont {Fossum}}]{Dom-16}%
  \BibitemOpen
  \bibfield  {author} {\bibinfo {author} {\bibfnamefont {P.}~\bibnamefont {Dommersnes}}, \bibinfo {author} {\bibfnamefont {A.}~\bibnamefont {Mikkelsen}},\ and\ \bibinfo {author} {\bibfnamefont {J.~O.}\ \bibnamefont {Fossum}},\ }\bibfield  {title} {\bibinfo {title} {Electro-hydrodynamic propulsion of counter-rotating pickering drops},\ }\href {https://doi.org/10.1140/epjst/e2016-60090-2} {\bibfield  {journal} {\bibinfo  {journal} {Eur. Phys. J. Spec. Top.}\ }\textbf {\bibinfo {volume} {225}},\ \bibinfo {pages} {699} (\bibinfo {year} {2016})}\BibitemShut {NoStop}%
\bibitem [{\citenamefont {Brosseau}\ \emph {et~al.}(2016)\citenamefont {Brosseau}, \citenamefont {Hickey},\ and\ \citenamefont {Vlahovska}}]{Bro-16}%
  \BibitemOpen
  \bibfield  {author} {\bibinfo {author} {\bibfnamefont {Q.}~\bibnamefont {Brosseau}}, \bibinfo {author} {\bibfnamefont {G.}~\bibnamefont {Hickey}},\ and\ \bibinfo {author} {\bibfnamefont {P.~M.}\ \bibnamefont {Vlahovska}},\ }\bibfield  {title} {\bibinfo {title} {{Electrohydrodynamic Quincke rotation of a prolate ellipsoid}},\ }\href@noop {} {\bibfield  {journal} {\bibinfo  {journal} {PNAS}\ } (\bibinfo {year} {2016})},\ \Eprint {https://arxiv.org/abs/1609.04384v1} {1609.04384v1} \BibitemShut {NoStop}%
\bibitem [{\citenamefont {Rozynek}\ \emph {et~al.}(2021)\citenamefont {Rozynek}, \citenamefont {Banaszak}, \citenamefont {Mikkelsen}, \citenamefont {Khobaib},\ and\ \citenamefont {Magdziarz}}]{Roz-21}%
  \BibitemOpen
  \bibfield  {author} {\bibinfo {author} {\bibfnamefont {Z.}~\bibnamefont {Rozynek}}, \bibinfo {author} {\bibfnamefont {J.}~\bibnamefont {Banaszak}}, \bibinfo {author} {\bibfnamefont {A.}~\bibnamefont {Mikkelsen}}, \bibinfo {author} {\bibfnamefont {K.}~\bibnamefont {Khobaib}},\ and\ \bibinfo {author} {\bibfnamefont {A.}~\bibnamefont {Magdziarz}},\ }\bibfield  {title} {\bibinfo {title} {Electrorotation of particle-coated droplets: from fundamentals to applications},\ }\href {https://doi.org/10.1039/D1SM00122A} {\bibfield  {journal} {\bibinfo  {journal} {Soft Matter}\ }\textbf {\bibinfo {volume} {17}},\ \bibinfo {pages} {4413} (\bibinfo {year} {2021})}\BibitemShut {NoStop}%
\bibitem [{\citenamefont {Melcher}\ and\ \citenamefont {Taylor}(1969)}]{Mel-69}%
  \BibitemOpen
  \bibfield  {author} {\bibinfo {author} {\bibfnamefont {J.}~\bibnamefont {Melcher}}\ and\ \bibinfo {author} {\bibfnamefont {G.}~\bibnamefont {Taylor}},\ }\bibfield  {title} {\bibinfo {title} {{Electrohydrodynamics: a review of the role of interfacial shear stresses}},\ }\href@noop {} {\bibfield  {journal} {\bibinfo  {journal} {Annual Review of Fluid Mechanics}\ }\textbf {\bibinfo {volume} {1}},\ \bibinfo {pages} {111} (\bibinfo {year} {1969})}\BibitemShut {NoStop}%
\bibitem [{\citenamefont {Washizu}\ and\ \citenamefont {Jones}(1994)}]{Was-94}%
  \BibitemOpen
  \bibfield  {author} {\bibinfo {author} {\bibfnamefont {M.}~\bibnamefont {Washizu}}\ and\ \bibinfo {author} {\bibfnamefont {T.}~\bibnamefont {Jones}},\ }\bibfield  {title} {\bibinfo {title} {Multipolar dielectrophoretic force calculation},\ }\href {https://doi.org/https://doi.org/10.1016/0304-3886(94)90053-1} {\bibfield  {journal} {\bibinfo  {journal} {Journal of Electrostatics}\ }\textbf {\bibinfo {volume} {33}},\ \bibinfo {pages} {187 } (\bibinfo {year} {1994})}\BibitemShut {NoStop}%
\bibitem [{\citenamefont {Premlata}\ and\ \citenamefont {WEI}(2019)}]{Pre-19}%
  \BibitemOpen
  \bibfield  {author} {\bibinfo {author} {\bibfnamefont {A.~R.}\ \bibnamefont {Premlata}}\ and\ \bibinfo {author} {\bibfnamefont {H.-H.}\ \bibnamefont {WEI}},\ }\bibfield  {title} {\bibinfo {title} {{The Basset problem with dynamic slip: slip-induced memory effect and slip{\textendash}stick~transition}},\ }\href@noop {} {\bibfield  {journal} {\bibinfo  {journal} {Journal of Fluid Mechanics}\ }\textbf {\bibinfo {volume} {866}},\ \bibinfo {pages} {431} (\bibinfo {year} {2019})}\BibitemShut {NoStop}%
\bibitem [{Note1()}]{Note1}%
  \BibitemOpen
  \bibinfo {note} {Even if $X(0)\protect \neq 0$, the additional term $X(0)M(s)$ goes to zero exponential as $s\protect \xrightarrow {}\infty $.}\BibitemShut {Stop}%
\bibitem [{\citenamefont {Sparrow}(1982)}]{Sparrowbook}%
  \BibitemOpen
  \bibfield  {author} {\bibinfo {author} {\bibfnamefont {C.}~\bibnamefont {Sparrow}},\ }\href@noop {} {\emph {\bibinfo {title} {The {L}orenz Equations: Bifurcations, Chaos, and Strange Attractors}}}\ (\bibinfo  {publisher} {Springer-Verlag, New York},\ \bibinfo {year} {1982})\BibitemShut {NoStop}%
\bibitem [{\citenamefont {Jackson}(1990)}]{jackson_1990}%
  \BibitemOpen
  \bibfield  {author} {\bibinfo {author} {\bibfnamefont {E.~A.}\ \bibnamefont {Jackson}},\ }\bibinfo {title} {Models based on third order differential systems},\ in\ \href {https://doi.org/10.1017/CBO9780511623981.003} {\emph {\bibinfo {booktitle} {Perspectives of Nonlinear Dynamics}}},\ Vol.~\bibinfo {volume} {2}\ (\bibinfo  {publisher} {Cambridge University Press},\ \bibinfo {year} {1990})\ p.\ \bibinfo {pages} {125–230}\BibitemShut {NoStop}%
\bibitem [{\citenamefont {Daitche}(2013)}]{Daitche2013}%
  \BibitemOpen
  \bibfield  {author} {\bibinfo {author} {\bibfnamefont {A.}~\bibnamefont {Daitche}},\ }\bibfield  {title} {\bibinfo {title} {Advection of inertial particles in the presence of the history force: Higher order numerical schemes},\ }\href {https://doi.org/10.1016/j.jcp.2013.07.024} {\bibfield  {journal} {\bibinfo  {journal} {J. Comput. Phys.}\ }\textbf {\bibinfo {volume} {254}},\ \bibinfo {pages} {93–106} (\bibinfo {year} {2013})}\BibitemShut {NoStop}%
\bibitem [{\citenamefont {Aguirre}\ \emph {et~al.}(2009)\citenamefont {Aguirre}, \citenamefont {Viana},\ and\ \citenamefont {Sanju\'an}}]{RevModPhys.81.333}%
  \BibitemOpen
  \bibfield  {author} {\bibinfo {author} {\bibfnamefont {J.}~\bibnamefont {Aguirre}}, \bibinfo {author} {\bibfnamefont {R.~L.}\ \bibnamefont {Viana}},\ and\ \bibinfo {author} {\bibfnamefont {M.~A.~F.}\ \bibnamefont {Sanju\'an}},\ }\bibfield  {title} {\bibinfo {title} {Fractal structures in nonlinear dynamics},\ }\href {https://doi.org/10.1103/RevModPhys.81.333} {\bibfield  {journal} {\bibinfo  {journal} {Rev. Mod. Phys.}\ }\textbf {\bibinfo {volume} {81}},\ \bibinfo {pages} {333} (\bibinfo {year} {2009})}\BibitemShut {NoStop}%
\bibitem [{\citenamefont {Pyragas}(2001)}]{Pyragas2001PRL}%
  \BibitemOpen
  \bibfield  {author} {\bibinfo {author} {\bibfnamefont {K.}~\bibnamefont {Pyragas}},\ }\bibfield  {title} {\bibinfo {title} {Control of chaos via an unstable delayed feedback controller},\ }\href {https://doi.org/10.1103/PhysRevLett.86.2265} {\bibfield  {journal} {\bibinfo  {journal} {Phys. Rev. Lett.}\ }\textbf {\bibinfo {volume} {86}},\ \bibinfo {pages} {2265} (\bibinfo {year} {2001})}\BibitemShut {NoStop}%
\bibitem [{\citenamefont {Pyragas}\ and\ \citenamefont {Pyragas}(2006)}]{Pyragas2006PRE}%
  \BibitemOpen
  \bibfield  {author} {\bibinfo {author} {\bibfnamefont {V.}~\bibnamefont {Pyragas}}\ and\ \bibinfo {author} {\bibfnamefont {K.}~\bibnamefont {Pyragas}},\ }\bibfield  {title} {\bibinfo {title} {Delayed feedback control of the lorenz system: An analytical treatment at a subcritical hopf bifurcation},\ }\href {https://doi.org/10.1103/PhysRevE.73.036215} {\bibfield  {journal} {\bibinfo  {journal} {Phys. Rev. E}\ }\textbf {\bibinfo {volume} {73}},\ \bibinfo {pages} {036215} (\bibinfo {year} {2006})}\BibitemShut {NoStop}%
\bibitem [{\citenamefont {Pyragas}(2006)}]{pyragas2006royal}%
  \BibitemOpen
  \bibfield  {author} {\bibinfo {author} {\bibfnamefont {K.}~\bibnamefont {Pyragas}},\ }\bibfield  {title} {\bibinfo {title} {Delayed feedback control of chaos},\ }\href {https://doi.org/10.1098/rsta.2006.1827} {\bibfield  {journal} {\bibinfo  {journal} {Philosophical Transactions of the Royal Society A: Mathematical, Physical and Engineering Sciences}\ }\textbf {\bibinfo {volume} {364}},\ \bibinfo {pages} {2309} (\bibinfo {year} {2006})}\BibitemShut {NoStop}%
\end{thebibliography}%

\end{document}